\documentclass[10pt,final,journal,twoside,twocolumn]{IEEEtran}

\usepackage{cite}
\ifCLASSINFOpdf
\usepackage[pdftex]{graphicx}
\else
\usepackage[dvips]{graphicx}
\fi

\usepackage[cmex10]{amsmath}
\usepackage{array}
\usepackage[caption=false,font=footnotesize]{subfig}
\usepackage{fixltx2e}
\usepackage{stfloats}
\usepackage{url}
\usepackage[mathlines]{lineno}
\usepackage{amsfonts,amssymb,amsthm}
\usepackage{cases}
\usepackage{psfrag}
\usepackage{setspace}
\usepackage{float}
\usepackage{color}

\newcommand{\bh}{\boldsymbol{h}}
\newcommand{\br}{\boldsymbol{r}}
\newcommand{\bw}{\boldsymbol{w}}
\newcommand{\bx}{\boldsymbol{x}}
\newcommand{\by}{\boldsymbol{y}}

\newcommand{\bI}{\boldsymbol{I}}
\newcommand{\bPi}{\boldsymbol{\Pi}}

\newcommand{\bbC}{\mathbb{C}}
\newcommand{\bbR}{\mathbb{R}}
\newcommand{\cB}{\mathcal{B}}

\newcommand{\cL}{\mathcal{L}}

\newcommand{\cO}{\mathcal{O}}
\newcommand{\cQ}{\mathcal{Q}}
\newcommand{\cR}{\mathcal{R}}
\newcommand{\cW}{\mathcal{W}}
\def\Rnn{{{\mathcal{R}^{nn}}}}
\def\Rnd{{{\mathcal{R}^{nd}}}}
\def\Rdn{{{\mathcal{R}^{dn}}}}
\def\Rdd{{{\mathcal{R}^{dd}}}}
\def\Bnn{{{\mathcal{B}^{nn}}}}
\def\Bnd{{{\mathcal{B}^{nd}}}}
\def\Bdn{{{\mathcal{B}^{dn}}}}
\def\Bdd{{{\mathcal{B}^{dd}}}}

\newcommand{\Rub}{\overline{R}}
\newcommand{\Rlb}{\underline{R}}
\newcommand{\xub}{\overline{x}}
\newcommand{\xlb}{\underline{x}}
\newcommand{\gamub}{\overline{\gamma}}
\newcommand{\gamlb}{\underline{\gamma}}

\newcommand{\alphat}{\tilde{\alpha}}
\newcommand{\betat}{\tilde{\beta}}

\newcommand{\st}{\tilde{s}}
\newcommand{\ut}{\tilde{u}}
\newcommand{\vt}{\tilde{v}}

\newcommand{\s}{^\star}
\newcommand{\PO}{^{\text{PO}}}
\newcommand{\ZF}{^{\text{ZF}}}
\newcommand{\MR}{^{\text{MR}}}

\newcommand{\nn}{^{nn}}

\newcommand{\dn}{^{dn}}
\newcommand{\dd}{^{dd}}

\newcommand{\norm}[1]{\left\lVert #1 \right\rVert}
\newcommand{\snorm}[1]{\left\lVert #1 \right\rVert^2}

\newtheorem{proposition}{Proposition}
\newtheorem{lemma}{Lemma}
\theoremstyle{definition}
\newtheorem{definition}{Definition}

\begin{document}

\title{Efficient Computation of Pareto Optimal Beamforming Vectors for the
  MISO Interference Channel with Successive Interference Cancellation}

\author{Johannes~Lindblom,~\IEEEmembership{Student Member,~IEEE,} 
Eleftherios~Karipidis,~\IEEEmembership{Member,~IEEE,} 
and~Erik~G.~Larsson,~\IEEEmembership{Senior~Member,~IEEE}%
\thanks{Manuscript received October 6, 2012; revised March 22, 2013 and May 31, 
2013; accepted June 10, 2013. This work has been supported in
  part by the Swedish Research Council (VR), the Swedish Foundation of
  Strategic Research (SSF), and the Excellence Center at
  Link\"oping-Lund in Information Technology (ELLIIT). This work has
  been performed in the framework of the European research project
  SAPHYRE, which was partly funded by the European Union under its FP7
  ICT Objective 1.1 - The Network of the Future. Preliminary
  versions of parts of the material in this paper were presented at
  ICASSP'11 \cite{Lindblom-2011-ICASSP} and CAMSAP'11 
  \cite{Lindblom-2011-CAMSAP}. }
\thanks{J. Lindblom and E. G. Larsson are with the Communication Systems 
Division, Department of Electrical Engineering (ISY), Link\"oping University, 
SE-581 83 Link\"oping, Sweden (e-mail: 
\{lindblom,erik.larsson\}@isy.liu.se).}%
\thanks{E. Karipidis was with the Communication Systems Division, Department 
of Electrical Engineering (ISY), Link\"oping University, SE-581 83 
Link\"oping, Sweden. He is now with Ericsson Research, Stockholm, Sweden 
(e-mail: karipidis@ieee.org).}}




\maketitle

\begin{abstract}
We study the two-user multiple-input single-output (MISO) Gaussian
interference channel where the transmitters have perfect channel state
information and employ single-stream beamforming. The receivers are
capable of performing successive interference cancellation, so when
the interfering signal is strong enough, it can be decoded, treating
the desired signal as noise, and subtracted from the received signal,
before the desired signal is decoded. We propose efficient methods to
compute the Pareto-optimal rate points and corresponding beamforming
vector pairs, by maximizing the rate of one link given the rate of the
other link. We do so by splitting the original problem into four
subproblems corresponding to the combinations of the receivers'
decoding strategies - either decode the interference or treat it as
additive noise. We utilize recently proposed parameterizations of the
optimal beamforming vectors to equivalently reformulate each
subproblem as a quasi-concave problem, which we solve very efficiently
either analytically or via scalar numerical optimization. The
computational complexity of the proposed methods is several
orders-of-magnitude less than the complexity of the state-of-the-art
methods. We use the proposed methods to illustrate the effect of the
strength and spatial correlation of the channels on the shape of the
rate region.  
\end{abstract}


\ifCLASSOPTIONpeerreview
\begin{center} \bfseries EDICS Category: MSP-CODR, MSP-CAPC, MSP-MULT, 
SAM-BEAM, SPC-INTF \end{center}
\fi
%
\IEEEpeerreviewmaketitle

\section{Introduction}

We study a wireless system where two adjacent transmitter (TX) --
receiver (RX) pairs, or links, operate simultaneously in the same
frequency band and interfere with each other. Each TX employs $n_T>1$
antennas, whereas each RX is equipped with a single antenna. Hence,
the system is modeled as the so-called multiple-input single-output
(MISO) interference channel (IC) \cite{Vishwanath-2004-ITW}. We assume
that the TXs have perfect knowledge of the local channels to both RXs
and use scalar Gaussian codes followed by single-stream
beamforming. Also, we assume that the RXs are capable to perform
successive interference cancellation (SIC)
\cite{Carleial-1978-TIT}. That is, when the interfering signal is
strong enough\footnote{In Sec.~\ref{sec:AchievableRegion}, we explain
  what mean by strong enough in the current context.}, a RX can decode
it and subtract it from the received signal before decoding the
desired signal. The decoding is done independently, since the RXs are
located apart and there is no coordination amongst them. SIC
capability is an important assumption because in principle it leads to
higher achievable rates than in case where the RXs treat interference
as noise. Note that the SIC rate region is a superset of the region
achieved when interference is treated as noise, since the latter is a
special case of the former. We see this by noting that the optimal
decoding process might be to directly decode the intended signal
treating the interference as noise. Once the intended signal is
decoded, the RXs are not interested in decoding the interference. The
resulting achievable rate region is defined by the so-called Pareto
boundary, which is the set of points where the rate of one link cannot
increase without decreasing the rate of the other.

The objective of this paper is to propose computationally efficient
methods for finding, in a centralized way, the Pareto-optimal (PO)
pairs of beamforming vectors which yield operating points on this
Pareto boundary. These methods are important because they enable a
fast computation of the rate region. Hence, we can use them to
illustrate how different channel realizations affect the shape of the
rate region or in the context of large-scale simulation studies of
interference networks. Moreover, they provide valuable insight on the
beamforming design and inspiration for practical implementations. 

The capacity region of the IC is in general unknown, but it is known
for certain cases. For strong interference, the capacity region
coincides with the capacity region for the case where both RXs decode
both messages, i.e., it is the intersection of the capacity regions of
two multiple-access channels, see \cite{Sato-1981-TIT, Han-1981-TIT,
  Costa-1985-TIT}. For weak interference it is optimal to treat it as
noise, see \cite{Motahari-2009-TIT} for the single-antenna IC and
\cite{Annapureddy-2011-TIT} for the multi-antenna IC. For the MISO IC,
various achievable rate regions have been proposed and studied, e.g.,
in \cite{Dahrouj-2011-TCOM, Ho-2010-Asilomar, Ho-2011-TIT,
  Jorswieck-2008-TSP, Mochaourab-2011-TSP}. Especially, the case of
treating interference as noise has been the subject of intense
studies, e.g., in \cite{Jorswieck-2008-TSP, Zakhour-2010-TWC,
  Zhang-2010-TSP, Mochaourab-2011-TSP, Shang-2011-TIT, Liu-2011-TSP,
  Mochaourab-2011-ICC,
  Mochaourab-2012-JSTSP}. In~\cite{Shang-2011-TIT}, it was shown that
single-stream beamforming is optimal for Gaussian codes. In
\cite{Jorswieck-2008-TSP}, a parameterization of the PO beamforming
vectors was proposed based on the properties that they use full
transmit power and lie in the subspace spanned by the local
channels. Alternative parameterizations were proposed in
\cite{Zakhour-2010-TWC} and \cite{Mochaourab-2011-TSP}, using the
concepts of virtual signal-to-interference-plus-noise ratio (SINR) and
gain regions, respectively. In \cite{Bjornson-2010-TSP}, the rate
region for the related scenario of cooperative multicell precoding was
characterized. The parameterizations for the two-user MISO IC
illustrate that for any number of transmit antennas, it can be reduced
to an equivalent MISO IC where each TX has two antennas
\cite{Annapureddy-2011-TIT}. 

The MISO IC with SIC-capable RXs was investigated in
\cite{Larsson-2008-JSAC} where the gain potential of SIC over treating
interference as noise was illustrated in terms of average rate at a
Nash equilibrium. Later, in  \cite{Ho-2010-Asilomar}, the achievable
rate region for SIC-capable RXs was formalized and in
\cite{Ho-2011-TIT} a parameterization for the PO beamforming vectors
was proposed, extending the respective result of
\cite{Jorswieck-2008-TSP}. A related work appears in
\cite{Dahrouj-2011-TCOM}, where a simplified version of the
Han-Kobayashi region \cite{Han-1981-TIT} was studied and semidefinite
relaxation was used to propose power control schemes and compute the
corresponding rate region. 

The parameterizations in \cite{Jorswieck-2008-TSP} and
\cite{Ho-2011-TIT} are useful analytical tools and they enable a
better intuitive understanding of the properties of the PO beamforming
vectors. When the RXs treat interference as noise, the PO beamforming
vectors are obtained by trading off between the conflicting objectives
of maximizing the desired signal power and minimizing the interference
\cite{Jorswieck-2008-TSP}. For SIC capable RXs, the trade-off is
between maximization of the desired signal power and maximization of
the interference, to enable SIC \cite{Ho-2011-TIT}. Another merit of
these parameterizations is that they substantially decrease the
dimension of the search space for a PO beamforming vector, from $n_T$
complex variables to one or two nonnegative real variables. However,
besides the dimensionality reduction, these parameterizations do not
directly provide a method for efficient computation of the Pareto
boundary. The reason is that they only constitute \emph{necessary}
conditions that the beamforming vector of each TX has to
\emph{separately} fulfill, whereas it is \emph{pairs} of beamforming
vectors that yield PO operating points. The state-of-the-art use of
the parameterizations has been to sample the parameters, consider all
possible combinations to generate a large number of achievable rate
pairs, and perform a brute-force search amongst them to find the ones
comprising the Pareto  boundary, e.g., see \cite{Jorswieck-2008-TSP}.

It is desirable to devise a method which directly and efficiently
computes PO points. Joint optimization of the beamforming vectors is
required for this purpose. As shown in \cite{Liu-2011-TSP}, the
problem of jointly maximizing a common utility function and of finding
PO points is NP-hard in general. Nevertheless, several methods have
been recently proposed, e.g., \cite{Zhang-2010-TSP} and
\cite{Karipidis-2010-WiOpt}, which apply successive convex
optimization techniques on the vector space of the beamforming vectors
to find the Pareto boundary when the RXs treat interference as
noise. The methods proposed herein achieve much higher computational
efficiency by optimizing instead on the parameter space which
characterizes the PO beamforming vectors of the SIC region. For the
case of treating interference as noise, this was attempted in
\cite{Jorswieck-2010-TCOM}, where monotonic optimization was used to
find specific PO points, e.g., the maximum sum-rate point. The method
proposed therein was faster than a brute-force search, but far less
efficient than the methods we propose. 

\subsection{Contributions and Organization}

In Sec. \ref{sec:systemmodel}, we give the system model. In
Sec. \ref{sec:AchievableRegion}, we define the SIC achievable rate
region and formulate the optimization problem that yields an arbitrary
point on the Pareto boundary, as a maximization of the rate of one
link given the rate achieved by the other link. In
Sec. \ref{sec:Rnn}--\ref{sec:Rdd}, we propose very efficient methods
to solve this problem, by combining, unifying, and improving the
preliminary approaches that appeared in our conference contributions
\cite{Lindblom-2011-ICASSP} and \cite{Lindblom-2011-CAMSAP}. The
common denominator of these methods is to exploit the
parameterizations and equivalently recast the maximization problem so
that it can be solved analytically or via scalar optimization. The
proposed method is applied on the following regions, whose union
constitutes the SIC region: 

1) In Sec. \ref{sec:Rnn}, we propose two methods to find the region
where both RXs treat the interference as noise. In the first method,
we equivalently formulate the originally non-concave rate maximization
problem as a scalar quasi-concave problem, which we solve optimally
with a gradient search approach. To find the Pareto boundary, we
repeat this optimization for various choices of the input rate. This
method is novel and improves the one in \cite{Karipidis-2010-WiOpt} in
two ways: a) we reduce the feasible set significantly from the set of
beamforming vectors to the parameter set defined in
\cite{Jorswieck-2008-TSP} and b) we solve a single quasi-concave
optimization problem instead of a sequence of convex feasibility
problems. In the second method, we use the KKT conditions to derive a
closed-form relation that couples the parameters of the TXs yielding a
PO pair. To find the Pareto boundary, we repeatedly solve the cubic
equation resulting from various choices for one of the
parameters. This method was presented in \cite{Lindblom-2011-ICASSP};
herein we give in addition a formal proof of global optimality of all
solutions to the KKT conditions. A similar result was independently
derived in \cite{Mochaourab-2011-ICC} and later extended in
\cite{Mochaourab-2012-JSTSP}. However, \cite{Mochaourab-2011-ICC} and
\cite{Mochaourab-2012-JSTSP} did not prove that all feasible solutions
to the corresponding cubic equation are global optima and potentially
they discard optimal solutions.

2) In Sec. \ref{sec:Rdn}, we find the two regions where one RX decodes
the interference, before decoding the desired signal, while the other
treats the interference as noise. We use the parameterization of
\cite{Ho-2010-Asilomar} to equivalently recast the rate maximization
problem as a quasi-concave problem of three real variables and
determine the solution in closed-form. This method was presented in
\cite{Lindblom-2011-CAMSAP}; herein, we give in addition a more
detailed derivation of the result. 

3) In Sec. \ref{sec:Rdd}, we find the region where both RXs decode the
interference. We use the parameterization of \cite{Ho-2010-Asilomar}
and split the rate maximization problem in two quasi-concave scalar
subproblems. This method improves the corresponding one in
\cite{Lindblom-2011-CAMSAP} in two ways: a) the number of variables is
decreased from four real variables to a single one and b) a single
instance of each of the two quasi-concave subproblems needs to be
solved instead of a sequence of convex feasibility problems. 

In Sec. \ref{sec:numerical}, we provide illustrations of the rate
regions for different channel properties and a complexity analysis. In
Sec.~\ref{sec:conclusion}, we make some concluding remarks. 

The source code that implements the proposed methods and generates the
illustrations in Sec.~\ref{sec:numerical} is available at
\url{http://urn.kb.se/resolve?urn=urn:nbn:se:liu:diva-93845}.

\subsection{Notation}

Boldface lowercase letters, e.g., $\bx$, denote column
vectors. $\{\cdot\}^H$ denotes the Hermitian (complex conjugate)
transpose of a vector. The Euclidean norm of a vector $\bx$ is denoted
$\norm{\bx}$. By $x \sim \mathcal{CN} (0,\sigma^2)$ we say that $x$ is
a zero-mean complex-symmetric Gaussian random variable with variance
$\sigma^2$. We denote the orthogonal projection onto the space spanned
by the vector $\bx$ as $\bPi_{\bx} \triangleq \bx
\bx^H/\snorm{\bx}$. The orthogonal projection onto the orthogonal
complement of $\bx$ is $\bPi_{\bx}^\perp \triangleq
\bI-\bPi_{\bx}$,where $\bI$ is the identity matrix. Note that for a
vector $\by$, we have $\snorm{\by} = \snorm{\bPi_{\bx} \by} +
\snorm{\bPi_{\bx}^\perp \by}$. We let $f'(x)$ and $f''(x)$ denote the
first and second derivatives, respectively, of a function $f(x)$. We
define $[x]^{\xub}_{\xlb} \triangleq \max\{\xlb, \min\{x,\xub\}\}$.

\section{System Model}
\label{sec:systemmodel}

We assume that the transmissions consist of scalar coding followed by 
single-stream (rank-1) beamforming and that all propagation channels are 
frequency-flat. The matched-filtered symbol-sampled complex baseband signal 
received by RX$_i$ is then modeled as
\begin{equation}
y_i = \bh_{ii}^H\bw_i s_i + \bh_{ji}^H\bw_j s_j + e_i, \quad  i,j \in 
\{1,2\}, \enspace j\neq i. \label{eq:model}
\end{equation}
In \eqref{eq:model}, $\bh_{ji} \in \bbC^{n_T}$ is the (conjugated) channel 
vector for the link $\text{TX}_j \rightarrow \text{RX}_i$. We assume that 
TX$_i$ perfectly knows the direct and crosstalk channels, $\bh_{ii}$ and 
$\bh_{ij}$, respectively, and that these are neither co-linear nor orthogonal. 
Also, $\bw_i \in\bbC^{n_T}$ is the beamforming vector employed by TX$_i$, $s_i 
\sim\mathcal{CN}(0,1)$ is the transmitted symbol of TX$_i$, and $e_i \sim 
\mathcal{CN} (0,\sigma_i^2)$ models the thermal noise at RX$_i$. The TXs have 
power constraints that we, without loss of generality, set to 1 and define the 
set of feasible beamforming vectors as $\cW \triangleq \{ \bw\in\bbC^{n_T} ~| 
\snorm{\bw} \leq 1 \}.$ The achievable rate for RX$_i$ depends on the received 
powers
\begin{equation} \label{eq:pq}
p_i(\bw_i) \triangleq |\bh_{ii}^H \bw_i|^2 \quad\text{and}\quad q_i(\bw_j)
\triangleq |\bh_{ji}^H \bw_j|^2
\end{equation}
over the direct and crosstalk channel, respectively.

In order to simplify the subsequent notation, we define the following 
channel-dependent constants. We define $g_{ij} \triangleq \norm{\bh_{ij}}$ and 
$\kappa_i \triangleq |\bh_{ij}^H \bh_{ii}| / (\norm{\bh_{ij}} 
\norm{\bh_{ii}})$, $j\neq i$. The latter is the cosine of the Hermitian angle 
between $\bh_{ii}$ and $\bh_{ij}$. When $\kappa_1=1$ or $\kappa_1 = 0$ the 
channels are parallel or orthogonal, respectively. Then, using these 
constants, we define 
\begin{align}
  \alpha_i & \triangleq \norm{\bPi_{\bh_{ij}} \bh_{ii}} = g_{ii}
  \kappa_i, \quad j\neq i, \label{eq:alpha} \\
  \tilde{\alpha}_i & \triangleq \norm{\bPi^\perp_{\bh_{ij}} \bh_{ii}}
  = \sqrt{g^2_{ii} - \alpha_i^2} = g_{ii} \sqrt{1-\kappa_i^2}, \quad
  j\neq i, \label{eq:alphatilde} \\
  \beta_i & \triangleq \norm{\bPi_{\bh_{ii}} \bh_{ij}} = g_{ij}
  \kappa_i, \quad j\neq i, \label{eq:beta} \\
  \tilde{\beta}_i & \triangleq \norm{\bPi^\perp_{\bh_{ii}} \bh_{ij}}
  = \sqrt{g^2_{ij} - \beta_i^2} = g_{ij} \sqrt{1-\kappa_i^2}, \quad
  j\neq i. \label{eq:betatilde}
\end{align}

\section{Achievable Rate Region of SIC Capable RXs}
\label{sec:AchievableRegion}
In this section, we give, for completeness, the definition of the achievable 
rate region for the described scenario \cite{Jorswieck-2008-TSP, Ho-2011-TIT}. 
We also denote the core optimization problem that we need to solve to find a 
point on the Pareto boundary.

Each pair of beamforming vectors $(\bw_1,\bw_2)$ and combination of decoding 
strategies (decode the interference ($d$) or treat it as noise ($n$)) is 
associated with a pair of maximum achievable rates. We denote by 
$R^{xy}_i(\bw_1,\bw_2)$ the rate of link~$i=1,2$, in bits per channel use 
(bpcu), where $x$ and $y$ are the decoding strategies ($n$ or $d$) of RX$_1$ 
and RX$_2$, respectively. For each pair of decoding strategies, we obtain a 
rate region by taking the union over all 1ible beamforming vectors, i.e.,
\begin{equation} \label{eq:RxyDef}
\cR^{xy} \triangleq \bigcup_{(\bw_1,\bw_2) \in \cW^2}
(R^{xy}_1(\bw_1,\bw_2), R^{xy}_2(\bw_1,\bw_2)).
\end{equation}

\textbf{The achievable rate region:}
The rate region for the MISO IC with SIC capability is obtained as the union 
of the regions corresponding to all decoding scenarios, i.e., $\cR = \cR^{nn} 
\cup \cR^{dn} \cup \cR^{nd} \cup \cR^{dd}.$ Next, for each decoding scenario 
and a given pair of beamforming vectors $(\bw_1, \bw_2)$, we give the maximum 
achievable rates \cite{Jorswieck-2008-TSP, Ho-2011-TIT}.

\textbf{$\cR^{nn}$ - Both RXs treat the interference as noise:} When both RXs 
treat the interference as noise, the maximum achievable rates of the links are 
\cite{Larsson-2008-JSAC}
\begin{equation} \begin{array}{l}
    R^{nn}_1(\bw_1,\bw_2) = \log_2 \left(1 + \frac{\displaystyle
        p_1(\bw_1)} {\displaystyle q_1(\bw_2) + \sigma_1^2} \right) 
    \quad \text{and} \\[1.1ex]
    R^{nn}_2(\bw_1,\bw_2) = \log_2 \left(1 + \frac{\displaystyle
	p_2(\bw_2)} {\displaystyle q_2(\bw_1) + \sigma_2^2}
    \right). \end{array}\label{eq:Rnn}
\end{equation}

\textbf{$\cR^{dn}$ - RX$_1$ decodes the interference, RX$_2$ treats it as 
additive noise:} Since RX$_1$ decodes and subtracts the interference caused by 
TX$_2$, it experiences an interference-free signal and the maximum achievable 
rate for link 1 is
\begin{equation} \label{eq:R1dn}
  R_1^{dn}(\bw_1) = \log_2 \left(1+ p_1(\bw_1) /\sigma_1^2 \right).
\end{equation}
RX$_1$ is able to decode the interference from TX$_2$, considering its own 
signal as noise, if the rate of link 2 is upper bounded by $\log_2 \left( 1 + 
q_1(\bw_2) / (p_1(\bw_1) + \sigma_1^2) \right).$ Since RX$_2$ does not decode 
the interference, the rate of link~2 is also upper bounded by $\log_2 \left(1 
+ p_2(\bw_2) / (q_2(\bw_1) + \sigma_2^2) \right).$ Hence the maximum 
achievable rate for link 2 is given by
\begin{align} 
  &R_2^{dn}(\bw_1, \bw_2) \nonumber \\
  & =  \log_2 \left( 1 + \min\left\{ \frac{\displaystyle q_1(\bw_2)}
      {\displaystyle p_1(\bw_1) + \sigma_1^2}, \frac{\displaystyle
        p_2(\bw_2)} {\displaystyle q_2(\bw_1) + \sigma_2^2} \right\}
  \right), \label{eq:R2dn} 
\end{align}
where we have used the fact that the logarithm is a monotonously increasing 
function.

For link 2, we note that the rate is not necessarily selected to fully utilize 
the signal-to-interference-plus-noise (SINR) ratio at RX$_2$. Actually, link 2 
might hold back on its rate to enable RX$_1$ to decode the interfering 
signal.\footnote{This fact was not exploited in \cite[Prop. 6 
a)]{Larsson-2008-JSAC}, so the description there led to an over-restrictive 
condition, hence, to a smaller achievable rate region.}

\textbf{$\cR^{nd}$ - RX$_2$ decodes the interference, RX$_1$ treats it as 
additive noise:} This case is identical to $\Rdn$, but with interchanged 
indices.

\textbf{$\cR^{dd}$ - Both RXs decode the interference:}
Both RXs decode the interference before decoding their desired signals. Since 
RX$_1$ decodes the interference from TX$_2$, the rate of link 1 is upper 
bounded by $ \log_2(1 + p_1(\bw_1) / \sigma_1^2).$ RX$_2$ can decode the 
interference caused by TX$_1$ if the rate of link 1 is upper bounded by $ 
\log_2 \left( 1 + q_2(\bw_1) / (p_2(\bw_2) + \sigma_2^2) \right).$ Then, the 
maximum achievable rate for link 1 is 
\begin{equation} \label{eq:R1dd}
  R_1^{dd}(\bw_1, \bw_2) =  \log_2 \left( 1 + \min\left\{
      \frac{p_1(\bw_1)} {\sigma_1^2}, \frac{\displaystyle q_2(\bw_1)}
      {\displaystyle p_2(\bw_2) + \sigma_2^2} \right\} \right). 
\end{equation}
By symmetry, the maximum achievable rate of link 2 is
\begin{equation} \label{eq:R2dd}
  R_2^{dd}(\bw_1, \bw_2) = \log_2 \left( 1 + \min\left\{ 
      \frac{p_2(\bw_2)}{\sigma_2^2} , \frac{\displaystyle q_1(\bw_2)}
      {\displaystyle p_1(\bw_1) + \sigma_1^2} \right\} \right).
\end{equation}

The problem of interest is to find the so-called Pareto boundary of the region 
$\cR$, which consists of PO rate pairs.
\begin{definition} \label{def:Pareto}
  A point $(R_1\s,R_2\s) \in \cR$ is (weakly) Pareto-optimal if there is no 
  other point $(R_1,R_2) \in \cR$ with $R_1 > R_1\s$ and $R_2 > R_2\s.$
\end{definition}

Graphically, the Pareto boundary is the north-east boundary of the region and 
due to Def. \ref{def:Pareto} it also includes the horizontal and vertical 
segments. In order to find the Pareto boundary of $\cR$, we first find the 
Pareto boundaries of $\Rnn,$ $\Rdn,$ $\Rnd,$ and $\Rdd$. Second, we consider 
as boundary of $\cR$ the boundary of the union of the four $\cR^{xy}$ regions. 
We denote by $\cB^{xy}$ the boundary of $\cR^{xy}$ and by $\cB$ the boundary of
$\cR$. In \cite[Lem. 1.2]{Bjornson-ResourceAllocation}, it is proven that 
$\Rnn$ is compact and normal under the assumptions in 
Sec.~\ref{sec:systemmodel}. It is straightforward to extend this proof to 
include $\Rnd,$ $\Rdn,$ and $\Rdd$ as well. Therefore, we conclude that the 
Pareto boundaries $\Bnn,$ $\Bdn,$ $\Bnd,$ and $\Bdd$ are closed.

We can find a point $(R_1\s,R_2\s)$ on $\cB^{xy}$ when the rate of one 
communication link, e.g., $R_1\s$, is given \cite[Prop. 
6.2]{Luenberger-MicroeconomicTheory}. The other rate, $R_2\s$, is the maximum 
one we simultaneously achieve, e.g., see the dashed lines in 
Fig.~\ref{fig:sym_HighCorr} for $\Rnn$, and we find it by the following rate 
optimization problem\footnote{Note that the optimization is only over the set 
of beamforming vectors satisfying the power constraints included in the set 
$\cW^2.$}
\begin{align}
	\underset{(\bw_1,\bw_2) \in \cW^2}{\text{maximize }} &
	R_2^{xy}(\bw_1,\bw_2) \label{eq:RoptObj} \\
	\text{subject to } & R_1^{xy}(\bw_1,\bw_2) = R_1\s. \label{eq:RoptCon}
\end{align}
The optimization \eqref{eq:RoptObj}--\eqref{eq:RoptCon} accepts as input the
coordinate $R_1\s$ of the sought PO rate pair and yields as optimal value the 
other coordinate $R_2\s$ and as optimal solution the enabling PO pair of 
beamforming vectors $(\bw_1\s, \bw_2\s).$ The choice of using $R_1\s$ as input 
to the optimization is arbitrary. By the symmetry of the problem, we can 
choose $R_2\s$ as input and have $R_1\s$ as the optimal value. In the next 
sections, we derive efficient methods for solving 
\eqref{eq:RoptObj}--\eqref{eq:RoptCon} for all SIC-constituent regions. Since 
the logarithm is monotonic, we use the equivalent reformulation of 
\eqref{eq:RoptObj}--\eqref{eq:RoptCon} as an SINR optimization, where the 
input parameter is $\gamma_1\s$, i.e., the SINR (or the SNR after interference 
cancellation) required to achieve $R_1\s$.

\section{Both RXs Treat the Interference as Noise}
\label{sec:Rnn}

In this section, we compute the boundary $\Bnn$. We let $\Rub\nn_1$ denote the 
maximum rate of link 1, achieved when TX$_1$ operates ``selfishly'' by using 
its maximum-ratio (MR) transmit beamforming vector $\bw_1\MR =
\arg\max_{\bw_1\in\cW} ~ p_1(\bw_1) = \bh_{11} / \norm{\bh_{11}} $ and
TX$_2$ operates ``altruistically'' by using its zero-forcing (ZF) transmit 
beamforming vector
$ \bw_2\ZF = \arg\max_{\substack{\bw_2\in\cW\\q_1(\bw_2)=0}} ~
p_2(\bw_2) = \bPi_{\bh_{21}}^\perp\bh_{22} /
\norm{\bPi_{\bh_{21}}^\perp \bh_{22}}$
\cite{Larsson-2008-JSAC}. This combination of transmit strategies yields the 
PO point $(\Rub\nn_1,\Rlb\nn_2) \triangleq \left( R_1\nn(\bw_1\MR,\bw_2\ZF), 
R_2\nn(\bw_1\MR,\bw_2\ZF)\right)$ where
\begin{align}
  \Rub\nn_1 \! &= \! \log_2\!\! \left(1 + \frac{\snorm{\bh_{11}}}
    {\sigma_1^2} \right) \!=\! \log_2\!\! \left(1 +
    \frac{g^2_{11}}{\sigma_1^2} \right) \text{ and} \label{eq:Rub_1nn}
  \\ 
  \Rlb\nn_2 \! &= \! \log_2
  \!\!\left(\!1+\frac{\snorm{\bPi_{\bh_{21}}^\perp \bh_{22}}} 
    {\snorm{\bPi_{\bh_{11}} \bh_{12}} + \sigma_2^2} \!\right)
  \! = \! \log_2 \!\!\left(\!1+\frac{\alpha_2^2}{\beta_1^2 + \sigma_2^2} 
    \!\right). \label{eq:Rlb_2nn}
\end{align}
The rate in \eqref{eq:Rlb_2nn} is the lowest strongly PO rate of link~2 
\cite{Larsson-2008-JSAC}. Interchanging the indices in 
\eqref{eq:Rub_1nn}--\eqref{eq:Rlb_2nn} we get the point
$(\Rlb\nn_1,\Rub\nn_2)$. As illustrated by the example in 
Fig.~\ref{fig:sym_HighCorr}, these points split $\Bnn$ into three segments.
The weakly PO horizontal (vertical) segment $[(0,\Rub\nn_2), 
(\Rlb\nn_1,\Rub\nn_2)]$ $\left([(\Rub\nn_1,0), (\Rub\nn_1,\Rlb\nn_2)]\right)$ 
is achieved when TX$_1$ (TX$_2$) uses the MR beamforming vector and TX$_2$ 
(TX$_1$) uses the ZF beamforming vector, adapting the transmit power in 
$[0,1]$.

The remainder of this section focuses on the strongly PO segment between 
$(\Rlb\nn_1,\Rub\nn_2)$ and $(\Rub\nn_1,\Rlb\nn_2)$. Inserting \eqref{eq:Rnn} 
into \eqref{eq:RoptObj}--\eqref{eq:RoptCon} and equivalently reformulating the 
rate maximization to SINR maximization, we obtain
\begin{align}
  & \underset{(\bw_1,\bw_2) \in \cW^2}{\text{maximize }} &&
  \frac{p_2(\bw_2)}{q_2(\bw_1) + \sigma_2^2} \label{eq:RnnObj} \\
  &\text{subject to } && \frac{p_1(\bw_1)}{q_1(\bw_2) + \sigma_1^2} = 
  \gamma_1\s. \label{eq:RnnCon}
\end{align}
From \eqref{eq:Rub_1nn}, we see that constraint \eqref{eq:RnnCon}, hence the 
optimization, is feasible when $\gamma_1\s \leq g_{11}^2/ \sigma_1^2$. The 
formulation \eqref{eq:RnnObj}--\eqref{eq:RnnCon} is non-convex since the 
objective function \eqref{eq:RnnObj} and the equality constraint 
\eqref{eq:RnnCon} consist of fractions of quadratics. 

In Secs. \ref{sec:WiOpt} and  \ref{sec:ICASSP}, we propose two methods to find 
very efficiently the global optimal solution. The main difference between 
these methods is the input required to yield the entire boundary; in the first 
method it is different choices for one of the PO SINR values, whereas in the 
second method it is different choices for one of the PO beamforming vectors. 
Both methods have computational complexity that is constant in the number of 
transmit antennas. The method in Sec.~\ref{sec:WiOpt} can be interpreted as 
solving an underlay cognitive radio problem where the secondary user, here 
link 2, maximizes its rate under various quality-of-service constraints for 
the primary user, here link 1. Also, this method can be used to determine if a 
rate point $(R_1,R_2)$ is feasible. Let $R_1\s = R_1$ be the input to  
\eqref{eq:RnnObj}--\eqref{eq:RnnCon}. Then, if $R_2\s \geq R_2,$ we can 
conclude that $(R_1,R_2)$ is feasible. The interpretation of the method in 
Sec.~\ref{sec:ICASSP} is that  TX$_1$ fixes its beamforming strategy and 
TX$_2$ seeks the best-response strategy to end up at a PO point.

\subsection{Numerical Method}
\label{sec:WiOpt}

The numerical method proposed in this section is a two-fold improvement of the 
one we presented in \cite{Karipidis-2010-WiOpt}. First, we exploit the 
parameterization \cite{Jorswieck-2008-TSP} of the PO beamforming vectors to 
reduce, without loss of optimality, the feasible set from $\cW^2$, i.e., a 
bounded convex set in $\mathbb{R}^{4n_T}$, to the bounded positive quadrant.
Second, we equivalently reformulate the optimization problem to further reduce 
the feasible set to a line segment in $\mathbb{R}$. The resulting problem is a 
scalar quasi-concave problem that needs to be solved once to yield a PO point, 
whereas the parameterized convex formulation in \cite{Karipidis-2010-WiOpt} 
required several iterations of the bisection method.

From \cite[Corollary 1]{Jorswieck-2008-TSP}, we know that the PO
beamforming vectors of the $\Rnn$ region can be parameterized as
\begin{equation} \label{eq:wnn}
  \bw_i (x_i) = x_i \frac{\bPi_{\bh_{ij}} \bh_{ii}}{\norm{\bPi_{\bh_{ij}}
      \bh_{ii}}} + \sqrt{1-x_i^2} \frac{\bPi_{\bh_{ij}}^\bot \bh_{ii}} 
  {\big\Vert\bPi_{\bh_{ij}}^\perp \bh_{ii}\big\Vert},
\end{equation}
where $0 \leq x_i \leq 1$ for $i,j=1,2$ and $j\neq i$. Note that
\eqref{eq:wnn} consists of a single nonnegative real parameter per
beamforming vector. Inserting \eqref{eq:wnn} into \eqref{eq:pq}, we get
\begin{align}
  p_i(\bw_i) &= \left( x_i \norm{\bPi_{\bh_{ij}}\bh_{ii}} + \sqrt{1-x_i^2} 
    \big\Vert\bPi_{\bh_{ij}}^\bot \bh_{ii}\big\Vert \right)^2 \nonumber \\
  & = \left( \alpha_i x_i + \tilde\alpha_i \sqrt{1-x_i^2} \right)^2, 
  \label{eq:pnn} \\
  q_j(\bw_i) & =  x_i^2 \frac{|\bh_{ij}^H \bh_{ii}|^2}
  {\snorm{\bPi_{\bh_{ij}}\bh_{ii}}} = g_{ij}^2 x_i^2. \label{eq:qnn}
\end{align}
From \cite{Mochaourab-2011-ICC}, we know that, for PO points, 
$x_i\leq\alpha_i/g_{ii} = \kappa_i$. This value maximizes \eqref{eq:pnn} and 
corresponds to the MR transmit beamforming vector. Further increase of $x_i$ 
will just increase the interference and decrease the desired signal power.

Inserting \eqref{eq:pnn} and \eqref{eq:qnn} into 
\eqref{eq:RnnObj}--\eqref{eq:RnnCon} and performing straightforward algebraic
manipulations, including taking the square root, we get the equivalent 
reformulation
\begin{align}
	& \underset{0 \leq x_1,x_2 \leq 1}{\text{maximize }} &&
	\frac{\alpha_2 x_2 + \tilde{\alpha}_2 \sqrt{1-x_2^2}} {\sqrt{g_{12}^2 
	x_1^2 + \sigma_2^2}} \label{eq:obj12} \\
	& \text{subject to } && \frac{\alpha_1 x_1 + \tilde{\alpha}_1 
	\sqrt{1-x_1^2}} {\sqrt{g_{21}^2 x_2^2 + \sigma_1^2}} = \sqrt{\gamma_1\s}, 
	\label{eq:con12}
\end{align}
where the constants $\alpha_1, \tilde\alpha_1, \alpha_2,
\tilde\alpha_2, g_{12}, g_{21}$ are all positive, as defined in Sec.
\ref{sec:systemmodel}. We further simplify the notation defining
\begin{align}
  u_i(x_i) & \triangleq \alpha_i x_i + \tilde{\alpha}_i \sqrt{1-x_i^2}
  \quad i=1,2 \quad\text{and}\label{eq:u_def} \\
  v_i(x_j) & \triangleq \sqrt{g_{ji}^2 x_j^2 + \sigma_i^2} \quad i,j =
  1,2, \enspace j\neq i. \label{eq:v_def}
\end{align}
It is straightforward to verify that $u_i(x_i)$ is concave and
non-decreasing for $x_i \leq \kappa_i$. Moreover, since $v_i(x_j)$ is
a norm, it is a convex and non-decreasing function in $x_j$. Then, we
make the following observation:
\begin{lemma} \label{lem:quasi-concave}
	The objective function \eqref{eq:obj12} and the left-hand-side (LHS)
	of the constraint \eqref{eq:con12} are quasi-concave functions.
\end{lemma}
\begin{IEEEproof}
  Note that $u_i(x_i)/v_i(x_j) \geq c$ is equivalent to $c v_i(x_j) -
  u_i(x_i) \leq 0$ since $v_i(x_i) > 0.$ Since $c v_i(x_j) $ is convex
  for all $c \geq 0$ and $u_i(x_i)$ is concave, $c v_i(x_j) - u_i(x_i)
  \leq 0$ defines a convex set. Hence, we conclude that the
  objective \eqref{eq:obj12} and the LHS of constraint \eqref{eq:con12} are 
  quasi-concave functions \cite[Ch. 3]{Boyd-ConvexOptimization}.
\end{IEEEproof}

Due to the equality in \eqref{eq:con12}, the problem 
\eqref{eq:obj12}--\eqref{eq:con12} is not quasi-concave as it stands
\cite{Arrow-1961-Econometrica}, but in the following we equivalently
reformulate it into a quasi-concave problem in one scalar variable. We solve 
equation \eqref{eq:con12} for $x_2$, keeping the positive root, as
\begin{equation} \label{eq:Rnn_x2}
x_2 =  \sqrt{\frac{u_1^2(x_1) - \gamma_1\s \sigma_1^2}{g_{21}^2 \gamma_1\s}} 
\triangleq w(x_1).
\end{equation}
Since a function of the form $\sqrt{t^2-a}$ is concave and
non-decreasing for $t \geq \sqrt{a},$ $a \geq 0$ and $u_1(x_1)$ is
concave and non-decreasing for $x_1 \leq \kappa_1,$ we conclude that
$w(x_1)$ is a concave and non-decreasing function of
$x_1 \leq \kappa_1$ \cite[Ch. 3.2]{Boyd-ConvexOptimization}.

The constraints $0 \leq x_2 \leq \kappa_2$ introduce lower and upper
bounds on $x_1$. Since $x_2 \geq 0$, it follows from \eqref{eq:Rnn_x2}
that $u_1(x_1) \geq \sqrt{\gamma_1\s \sigma_1^2}$. Then, from
\eqref{eq:u_def} and $x_1 \geq 0,$ it follows that we must have $x_1 \geq 
\xlb_1$, where
\begin{equation} \label{eq:xlb}
  \xlb_1 \triangleq \max\left\{0, \kappa_1\sqrt{ \frac{\gamma_1\s} 
      {\gamub_1^{nn}}} - 
    \sqrt{1-\kappa_1^2} \sqrt{1-\frac{\gamma_1\s} {\gamub_1^{nn}}} \right\}.
\end{equation}
Note that $\xlb_1$ is real for $\gamma_1\s \leq \gamub_1^{nn} \triangleq 
2^{\Rub\nn_1}-1 = g_{11}^2/\sigma_1^2$. Furthermore, for the upper limits we 
must have $x_2 = w(x_1) \leq \kappa_2$ and $x_1 \leq \kappa_1,$ which
imply $x_1 \leq \xub_1$, where 
\begin{equation}
  \hspace*{-2mm} \xub_1 \triangleq \left\{ \begin{array}{ll}
      \kappa_1 \sqrt{\frac{ \displaystyle \gamma_1\s} { \displaystyle 
          \gamma_1\MR}} - \sqrt{1-\kappa_1^2} 
      \sqrt{1-\frac{\displaystyle \gamma_1\s} {\displaystyle \gamma\MR}} , & 
      \!\!\! \gamma_1\s \leq \gamma_1\MR, \\
      \kappa_1, & \!\!\! \gamma_1\s > \gamma_1\MR,
    \end{array}  \right.\!\!  \label{eq:xub}
\end{equation}
and where $\gamma_1\MR \triangleq g_{11}^2 /(g_{21}^2 \kappa_1^2 + 
\sigma_1^2)$ is the SINR of link 1 when both TXs use the MR beamforming 
vectors, which yield the so-called Nash Equilibrium \cite{Larsson-2008-JSAC}. 
Note that $\gamma_1\MR < \gamub\nn_1,$ where the latter is the SNR of RX$_1$ 
at the single-user point of the rate region. It can be verified that $\xub_1 
\geq 0$ since $\gamma\s_1 \geq \gamlb\nn_1 \triangleq 2^{\Rlb\nn_1}-1.$

Inserting \eqref{eq:Rnn_x2} in \eqref{eq:obj12}, along with the lower and 
upper bounds \eqref{eq:xlb} and \eqref{eq:xub}, respectively, yields the 
scalar optimization problem
\begin{equation} \label{eq:Rnn_scalar_opt}
  \underset{\xlb_1 \leq x_1 \leq \xub_1}{\text{maximize }} \quad
  \frac{ u_2\left( w(x_1)\right)} {v_2(x_1)} \triangleq s(x_1).
\end{equation}
Note that the objective function corresponds to the square root of the SINR of 
link 2, i.e, $s(x_1) = \sqrt{\gamma_2}$. Next, we study its properties and 
prove that it is quasi-concave.
\begin{lemma} \label{lem:stationary}
	The function $s(x_1\!)$ is quasi-concave for $\xlb_1 < x_1 < \xub_1.$ 
\end{lemma}
\begin{IEEEproof}
  First, we observe that $s(x_1)$ is at least twice continuously
  differentiable for $\xlb_1 < x < \xub_1$. Second, we show that
  $s'(x_1) = 0$ implies that $s''(x_1) < 0$ and it follows that
  $s(x_1)$ is quasi-concave
  \cite[Ch. 3.4.3]{Boyd-ConvexOptimization}. 
	
  The first derivative of $s(x_1)$ is
  \begin{equation} \label{eq:Rnn_FirstDerivative}
    s'(x_1) = \frac{w' (x_1) u_2'(w (x_1)) v_2 (x_1) -
      u_2( w (x_1)) v_2'(x_1)}{ v_2^2 (x_1) }
  \end{equation}
  and the second derivative is
  \begin{align}
    & s''(x_1) = \frac{1}{v_2^2 (x_1)} \big( w''(x_1) u_2'(w(x_1)) 
    v_2(x_1) + \nonumber \\
    &+ (w'(x_1))^2 u_2''(w (x_1)) v_2(x_1) - u_2( w (x_1))
    v_2''(x_1)\big).  \label{eq:Rnn_SecondDerivative}
  \end{align}
  We know that $w(x_1)$ is concave and non-decreasing for $x_1 \leq
  \kappa_1$, $v_2(x_1)$ is convex and non-decreasing and $u_2(x_2)$ is
  concave.\footnote{We could have used the fact the $u_2(w(x_1))$ is
    also non-decreasing for $x_1 \leq \xub_1,$ to obtain a simpler
    proof. However, in Sec \ref{sec:Rdd}, we need this more general
    case.} Therefore, we conclude that the second and third terms of
  \eqref{eq:Rnn_SecondDerivative} are non-positive. Also, we note that
  the second term is zero only if $x_1 = \kappa_1$ and the third term
  is zero only if $x_1 = 0$. Hence, we conclude that the sum of the
  second and third terms in \eqref{eq:Rnn_SecondDerivative} is always
  negative. It remains to show that the first term in
  \eqref{eq:Rnn_SecondDerivative} is non-positive for a stationary
  point $x_1\s$. We know that $w''(x_1\s) \leq 0$ and $v_2(x_1\s) >
  0$, so we must show that $u_2'(w(x_1\s)) \geq 0$. For a stationary
  point, the first derivative is zero, so from
  \eqref{eq:Rnn_FirstDerivative} it follows that 
  \begin{equation} \label{eq:u2_prim_sign}
    u_2'(w (x_1\s)) = \frac{u_2( w (x_1\s)) v_2'(x_1\s)} {w' (x_1\s) v_2 
      (x_1\s)} \geq 0
  \end{equation}
  since $u_2( w (x_1)),$ $v_2'(x_1),$ $w' (x_1)$, and $v_2 (x_1)$ all
  are non-negative for $x_1 \leq \kappa_1.$  Since $s'(x_1) = 0$
  implies $s''(x_1) < 0$, we conclude that $s(x_1)$ is quasi-concave,
  \cite[Ch. 3.4.3]{Boyd-ConvexOptimization} 
\end{IEEEproof}

Since problem \eqref{eq:Rnn_scalar_opt} has a single real variable and the 
objective function is quasi-concave, the optimum solution can be found very 
efficiently. Since the objective function is monotonously increasing 
(decreasing) to the left (right) of the stationary point, a gradient method 
can be used. In Tab. \ref{tab:Rnn_num}, we propose a method that computes 
strongly PO points of $\Rnn$. As input, the method requires the channel 
constants, the noise variances, and the number $M$ of requested boundary 
points. The output is stored in the vectors $\br_1, \br_2 \in \bbR^M$. The 
rates in $\br_1$ are obtained by uniform sampling over the interval 
$[\Rlb\nn_1, \Rub\nn_1]$. In line 4, we compute the end point $(\Rlb\nn_1, 
\Rub\nn_2)$. For each boundary point, we compute the lower and upper bounds 
$\xlb_1$ and $\xub_1$, respectively, and ensure that the solution lies in the 
interval $[\xlb_1,\xub_1]$. In line 8, the $x_1\s$ corresponding to the 
previously computed point is used as initial value for the next point on the 
boundary. The reason is that we expect that the solution will not change 
significantly for two nearby points. In lines 10--14, we find the optimal 
solution $x_1\s$ by a gradient ascend method. In each repetition, we compute 
the derivative $s'(x_1)$ and then find a step length $t$, by backtracking line 
search \cite[Ch. 9.2]{Boyd-ConvexOptimization}. This is repeated until the 
improvement from the previous iteration is smaller than some predefined 
tolerance $\epsilon$. Since $s (x_1)$ is quasi-concave, its derivative can be 
small without being close to the optimum. Hence, $\epsilon$ has to be chosen 
very small. In line 20, the end point $(\Rub\nn_1,\Rlb\nn_2)$ is computed.

\begin{table}[ht]
  \begin{center}
    \begin{tabular}{|rl|}
      \hline
      1: & \textbf{Input: } $g_{ij},$ $\kappa_i,$ $\sigma_i^2,$ 
      $i,j=1,2,$ $M,$
      and $\epsilon$ \\
      2: & \textbf{Output: } $\Bnn$ given by vectors $\br_1,
      \br_2 \in \bbR^M$ \\
      3: & $\br_1 = \left[\Rlb\nn_1: (\Rub\nn_1 - \Rlb\nn_1)/(M-1): 
        \Rub\nn_1\right]$ \\
      4: & $\br_2(1) = \Rub\nn_2,$ $x_1\s = 0$ \\
      5: & \textbf{for } $k = 2:M-1$ \\
      6: & ~~~ $\gamma_1\s = 2^{\br_1(k)}-1$ \\
      7: & ~~~ Compute $\xlb_1$ and $\xub_1$ using \eqref{eq:xlb} and 
      \eqref{eq:xub} \\
      8: & ~~~ $x_1^{(0)} = [x_1\s]_{\xlb_1}^{\xub_1}$ \\
      9: & ~~~ $l = 0$ \\
      10: & ~~~ \textbf{repeat } \\
      11: & ~~~~~~ Compute $s'(x_1^{(l)})$ and determine step size $t$
      \\
      12: & ~~~~~~ $x_1^{(l+1)} = [x_1^{(l)}+t s'(x_1^{(l)})]_{\xlb_1}^{\xub_1}$\\
      13: & ~~~~~~ $l \leftarrow l+1$ \\
      14: & ~~~ \textbf{until } $\big|s(x_1^{(l)})-s(x_1^{(l-1)})\big| < 
      \epsilon$ \\
      15: & ~~~ $x_1\s = x_1^{(l)}$ \\
      16: & ~~~ Compute $x_2\s = w(x_1\s)$ using \eqref{eq:Rnn_x2} \\
      17: & ~~~ Compute $\bw_i\s = \bw_i(x_i\s)$ using \eqref{eq:wnn} \\
      18: & ~~~ Compute $\br_2(k) = R_2\s = R^{nn}_2(\bw_1\s,\bw_2\s)$
      using \eqref{eq:Rnn} \\ 
      19: & \textbf{end} \\
      20: & $\br_2(M) = \Rlb\nn_2,$ $x_1\s = \kappa_1,$ $x_2\s=0$ \\
      \hline
    \end{tabular}
  \end{center}
  \caption{Numerical method to compute $\Bnn$}
  \label{tab:Rnn_num}
\end{table}

\subsection{Closed-Form Parameterization} \label{sec:ICASSP}

In this section, we use the Karush-Kuhn-Tucker (KKT) conditions of the 
optimization problem \eqref{eq:obj12}--\eqref{eq:con12} in order to derive a 
closed-form relation between the parameters of the beamforming vectors that 
jointly yield a PO rate point. A preliminary version of this method was 
presented in \cite{Lindblom-2011-ICASSP}; herein, we elaborate the derivations 
and provide a proof of global optimality. The latter is achieved using the 
parameterization \eqref{eq:wnn}, whereas a different parameterization was used 
in \cite{Lindblom-2011-ICASSP}.

In general, the KKT conditions only provide necessary conditions for global 
optimality. However, we show that for this specific problem, the KKT 
conditions are also sufficient. Towards this direction, we relax the equality 
constraint \eqref{eq:con12} to a lower-bound inequality.\footnote{By 
contradiction, we can show that this relaxation is tight at the optimum. 
Assume that the optimal solution meets \eqref{eq:con12} by strict inequality. 
Then there is room to increase $x_2$ or decrease $x_1$ in order to make the 
objective \eqref{eq:obj12} larger. This is illustrated by the dashed lines in 
Fig. \ref{fig:sym_HighCorr}. Hence the relaxed problem is equivalent to the 
original one.} Then, due to Lem. \ref{lem:quasi-concave}, the relaxed 
optimization problem \eqref{eq:obj12}--\eqref{eq:con12} falls into the class 
of quasi-concave problems \cite{Arrow-1961-Econometrica}.  Th. 1 in 
\cite{Arrow-1961-Econometrica} gives a number of sufficient conditions for 
global optimality of the solution to the KKT conditions of a constrained 
quasi-concave program. It suffices that one of these conditions is satisfied. 
Condition a) is that the gradient of the objective function should have at 
least one negative component for a solution that satisfies the KKT conditions. 
By simple inspection of the objective function \eqref{eq:obj12}, it follows 
that:
\begin{lemma} \label{lem:non-decreasing}
  The objective \eqref{eq:obj12} is decreasing with $x_1 \geq  0$, for
  fixed $x_2$.
\end{lemma}
Hence, due to Lem. \ref{lem:non-decreasing}, the relaxed version of the 
problem \eqref{eq:obj12}--\eqref{eq:con12} satisfies condition a) of Th. 1 in 
\cite{Arrow-1961-Econometrica}. Then, from Lem. \ref{lem:quasi-concave}, Lem. 
\ref{lem:non-decreasing}, and \cite[Th. 1]{Arrow-1961-Econometrica}, we have 
the following result:
\begin{proposition} \label{prop:Suff_conditions}
  The KKT conditions of the relaxed problem
  \eqref{eq:obj12}--\eqref{eq:con12} are  sufficient conditions for
  global optimality.
\end{proposition}

For notational convenience, we make the bounding constraints on $x_i$ 
implicit, i.e., we declare a solution of the KKT conditions feasible only if 
it adheres to the bounding constraints. The Lagrange function of the relaxed 
\eqref{eq:obj12}--\eqref{eq:con12} is
\begin{equation} \label{eq:lagrange}
  \cL(x_1, x_2, \mu) = \frac{u_2(x_2)} {v_2(x_1)} + \mu \left(
    \frac{u_1(x_1)} {v_1(x_2)} - \sqrt{\gamma_1\s} \right),
\end{equation}
where the Lagrange multiplier $\mu$ is non-negative. Hence, the KKT
conditions are \cite{Boyd-ConvexOptimization}
\begin{align}
  \mu \left( \frac{u_1(x_1)} {v_1(x_2)} - \sqrt{\gamma_1\s} \right) & = 0, 
  \label{eq:KKT_SINR_comp_slack} \\
  \frac{\partial \cL}{\partial x_1} = -\frac{v_2'(x_1) u_2(x_2)} {v_2^2(x_1)}
  + \mu \frac{u_1'(x_1)} {v_1(x_2)}  & = 0, \label{eq:KKT_x1} \\
  \frac{\partial \cL}{\partial x_2} = \frac{u_2'(x_2)} {v_2(x_1)} -
  \mu \frac{v_1'(x_2) u_1(x_1)} {v_1^2(x_2)} & = 0. \label{eq:KKT_x2}
\end{align}
In \eqref{eq:KKT_SINR_comp_slack}--\eqref{eq:KKT_x2}, we avoided
explicitly including the primal feasibility constraints, since
\eqref{eq:obj12}--\eqref{eq:con12} is always feasible if we choose
$\gamma_1\s \leq g_{11}^2/\sigma_1^2$. Also, it is straightforward to verify 
that the corresponding Lagrange multipliers must be zero. We use the KKT 
conditions to find a relation between the parameters $x_1$ and $x_2$ that 
jointly yield PO points. Since we are not looking for a specific PO point, we 
can disregard condition \eqref{eq:KKT_SINR_comp_slack}. Once we have found a 
pair $(x_1\s,x_2\s)$ that solves \eqref{eq:KKT_x1}--\eqref{eq:KKT_x2} for some 
$\mu\s$, we insert the triplet $(x_1\s,x_2\s,\mu\s)$ into 
\eqref{eq:KKT_SINR_comp_slack} to find $\gamma_1\s.$ Clearly 
$(x_1\s,x_2\s,\gamma_1\s,\mu\s)$ solves the KKT conditions 
\eqref{eq:KKT_SINR_comp_slack}--\eqref{eq:KKT_x2}.

When $x_1 > 0$, we can verify from \eqref{eq:KKT_x1} that $\mu > 0$. Then, we
use \eqref{eq:KKT_x1} and \eqref{eq:KKT_x2} to solve for $\mu$. Equating the 
solutions, we get the relation
\begin{equation} \label{eq:xy_relation}
\frac{u_2'(x_2) v_1^2(x_2)} {v_1'(x_2) v_2(x_1) u_1(x_1)} =
\frac{v_2'(x_1) v_1(x_2) u_2(x_2)} {u_1'(x_1) v_2^2(x_1)}.
\end{equation}
By collecting all functions of $x_1$ and $x_2$ in the LHS and
RHS, respectively, we equivalently rewrite \eqref{eq:xy_relation} as
\begin{equation} \label{eq:xy_relation_rw}
  f(x_1) \!\triangleq \!\frac{u_1'(x_1) v_2^2(x_1)} {v_2'(x_1) v_2(x_1)
    u_1(x_1)} \!=\! \frac{v_1'(x_2) v_1(x_2) u_2(x_2)} {u_2'(x_2)
    v_1^2(x_2)} \!\triangleq\! g(x_2).
\end{equation}
The LHS and RHS of \eqref{eq:xy_relation_rw} are functions of only $x_1$ and 
$x_2$, respectively, which we denote as $f(x_1)$ and $g(x_2)$, respectively. 
In order to find a PO point, we fix $x_1$ at a specific value $x_1\s$ and then
solve $g(x_2)=f(x_1\s)$ to get $x_2\s$.

Due to the square roots, it is complicated to solve for $x_2$ as it stands. 
Instead, we use the alternative parameterization, that the PO beamforming 
vectors are linear combinations of the MR and ZF beamforming vectors 
\cite[Corollary 2]{Jorswieck-2008-TSP}, i.e.,
\begin{equation}\label{eq:w_lambda}
\bw_i\PO(\lambda_i) = \frac{\lambda_i \bw_i\MR + (1-\lambda_i) \bw_i\ZF}
{\norm{\lambda_i \bw_i\MR + (1-\lambda_i) \bw_i\ZF}},
\end{equation}
where $\lambda_i \in [0,1].$ To go from the parameterization in
\eqref{eq:wnn} to that in \eqref{eq:w_lambda}, we use the mapping
\begin{align}
  x_i & = \phi_i(\lambda_i) \triangleq 
  \frac{\kappa_i \lambda_i} {\norm{\lambda_i \bw_i\MR + (1-\lambda_i)
      \bw_i\ZF}} \nonumber \\
  & = \frac{\kappa_i \lambda_i} {\sqrt{ 2
      \rho_i \lambda_i^2 - 2 \rho_i \lambda_i + 1}}, 
  \label{eq:x_lambda_mapping}
\end{align}
where $\rho_i \triangleq
1-\sqrt{1-\kappa_i^2}.$ Since $0 < \kappa_i < 1$, we have
\begin{equation} \label{eq:mapping_deriv}
  \frac{d \phi_i} {d \lambda_i} = \frac{\kappa_i(1-\rho_i \lambda_i)} {( 2
    \rho \lambda_i^2 - 2 \rho_i \lambda_i + 1)^{3/2}} > 0.
\end{equation}
Hence, it follows that \eqref{eq:x_lambda_mapping} is a one-to-one mapping 
between $x_i$ and $\lambda_i$ and the problem of solving $g(x_2) = 
f(x_1\s)$ with respect to $x_2$ is equivalent to that of solving 
$g(\phi_2(\lambda_2)) = f(\phi_1(\lambda_1\s))$ with respect to $\lambda_2$. 
By inserting \eqref{eq:x_lambda_mapping} into \eqref{eq:xy_relation_rw}, we 
equivalently write $g(\phi_2(\lambda_2))=f(\phi_1(\lambda_1\s))$ as
\begin{equation} \label{eq:lambda_rel}
  \frac{\lambda_2 (1-\rho_2 \lambda_2) (\rho_2 \lambda_2 +
    (1-\rho_2))} {(1-\lambda_2) (\rho_2(2-\rho_2+2\zeta_2)
    \lambda_2^2 - 2\rho_2\zeta_2 \lambda_2 + \zeta_2)}=f(\phi_1(\lambda_1\s)),
\end{equation}
where $\zeta_i \triangleq \sigma^2_j / g_{ij}^2, \quad i,j=1,2, \quad j\neq i.$
In \eqref{eq:lambda_rel}, we see that $g(\phi_2(\lambda_2))$ is a
fraction of cubic polynomials. Since $f(\phi_1(\lambda_1\s))$ is a
constant, we write \eqref{eq:lambda_rel} as the cubic equation
\begin{equation} \label{eq:cubic_eq}
  c_3 \lambda_2^3 + c_2 \lambda_2^2  + c_1 \lambda_2 + c_0 = 0.
\end{equation}
The coefficients of the cubic equation \eqref{eq:cubic_eq} are
\begin{equation} \label{eq:coefficients}
  \left\{ \begin{array} {l}
      c_0 \triangleq -\zeta_2  f(\lambda_1\s), \\
      c_1 \triangleq (1+2\rho_2) \zeta_2 f(\lambda_1\s)+ (1-\rho_2), \\
      c_2 \triangleq -\rho_2 (2-\rho_2+4 \zeta_2) f(\phi_1(\lambda_1\s)) + 
      \rho_2^2, \\
      c_3 \triangleq \rho_2(2 - \rho_2 + 2 \zeta_2)
      f(\phi_1(\lambda_1\s)) - \rho_2^2. 
    \end{array} \right.
\end{equation}
Cubic equations can be solved in closed form
\cite{Abramowitz-Handbook}. The roots of \eqref{eq:cubic_eq} are three
candidates for $\lambda_2\s$.\footnote{Note that any other choice of
  $\lambda_1\s$ gives us a point in the interior of $\Rnn$.} Since
$\lambda_1\s\in[0,1]$, we have the following three cases.
\begin{itemize}
\item $\lambda_1\s = 0$: From  \eqref{eq:Rlb_2nn} and \eqref{eq:w_lambda}
  we know that $\lambda_2\s=1$.
  
\item $0 < \lambda_1\s < 1$: We find the roots of \eqref{eq:cubic_eq}
  and keep the roots that satisfy the constraint $0 \leq \lambda_2
  \leq 1$. We can potentially have more than one feasible root, but
  from Prop.~\ref{prop:Suff_conditions}, we know that all feasible
  roots yield a PO solution.\footnote{In \cite{Mochaourab-2011-ICC}
    and \cite{Mochaourab-2012-JSTSP}, it was not made clear whether
    all feasible solutions to the corresponding cubic equation are
    optimal or not. Especially, equation (25) in
    \cite{Mochaourab-2011-ICC} provides only necessary conditions for
    Pareto optimality \cite[Ch. 6]{Luenberger-MicroeconomicTheory}.}
	
\item $\lambda_1\s = 1$: Again, from \eqref{eq:Rlb_2nn} and
  \eqref{eq:w_lambda} we see that $\lambda_2\s=0$.
\end{itemize}

The overall method to compute the entire $\Bnn$ is summarized in
Tab. \ref{tab:Rnn_algorithm}. By uniform sampling of $\lambda_1\s$ over the 
interval $[0,1]$, we will cover the entire interval $[\Rlb^{nn},\Rub^{nn}].$ 
Once we have found the coefficients in Sec. \ref{sec:systemmodel}, the 
complexity  is constant in the number of antennas.

\begin{table}[H]
  \begin{center}
    \begin{tabular}{|rl|}
      \hline
      1: & \textbf{Input and output:} same as in Tab. \ref{tab:Rnn_num} 
      \\
      2: & \textbf{for } $\lambda_1\s = [0:1/(M-1):1]$ \\
      3: & ~~~ Compute $f(\phi_1(\lambda_1\s))$ according to
      \eqref{eq:xy_relation_rw} and \eqref{eq:x_lambda_mapping}\\
      4: & ~~~ $\lambda_2\s = \text{roots of cubic equation
        \eqref{eq:cubic_eq} that are in } [0,1]$ \\
      5: & ~~~ Compute the rate point(s) using \eqref{eq:Rnn},
      \eqref{eq:wnn}, and \eqref{eq:x_lambda_mapping} \\
      6: & \textbf{end} \\
      \hline
    \end{tabular}
  \end{center}
  \caption{Closed-form method to compute $\Bnn$}
  \label{tab:Rnn_algorithm}
\end{table}

\section{Only one RX Decodes the Interference}
\label{sec:Rdn}

In this section, we compute $\Bdn$ on closed form. A
condensed description of this method was given in \cite{Lindblom-2011-CAMSAP}.

Inserting the rate expressions \eqref{eq:R1dn} and \eqref{eq:R2dn} in
the optimization problem \eqref{eq:RoptObj}--\eqref{eq:RoptCon}
and equivalently reformulating the rate maximization to SINR maximization, we 
obtain
\begin{align}
  & \underset{(\bw_1,\bw_2) \in
    \cW^2}{\text{maximize }} && \min \left\{ \frac{q_1(\bw_2)}
    {p_1(\bw_1) + \sigma_1^2},
    \frac{p_2(\bw_2)} {q_2(\bw_1) + \sigma_2^2} \right\}
  \label{eq:RdnObj} \\
  &\text{subject to } && p_1(\bw_1)/\sigma_1^2 = \gamma_1\s.
  \label{eq:RdnCon1}
\end{align}
As with \eqref{eq:RnnObj}--\eqref{eq:RnnCon}, we see that 
\eqref{eq:RdnObj}--\eqref{eq:RdnCon1} is feasible when $\gamma_1\s \leq 
g_{11}^2/ \sigma_1^2$. The formulation \eqref{eq:RdnObj}--\eqref{eq:RdnCon1} 
is nonconvex, because the objective \eqref{eq:RdnObj} is the minimum of two 
fractions of quadratic functions and \eqref{eq:RdnCon1} is a quadratic
equality constraint.

In \cite{Ho-2011-TIT}, it was shown that the PO beamforming vectors of the 
$\Rdn$ region can be parameterized as
\begin{align}
  \bw_1(x_1,y_1) & = x_1 \frac{\bPi_{\bh_{12}} \bh_{11}}
  {\norm{\bPi_{\bh_{12}} \bh_{11}}} + y_1 \frac{\bPi_{\bh_{12}}^\bot
    \bh_{11}} {\big\Vert\bPi_{\bh_{12}}^\bot
    \bh_{11}\big\Vert}, \label{eq:w1dn} \\
  \bw_2(x_2) & = x_2 \frac{\bPi_{\bh_{22}} \bh_{21}}
  {\norm{\bPi_{\bh_{22}} \bh_{21}}} + \sqrt{1-x_2^2}
  \frac{\bPi_{\bh_{22}}^\bot \bh_{21}} {\big\Vert\bPi_{\bh_{22}}^\bot
    \bh_{21}\big\Vert}, \label{eq:w2dn} 
\end{align}
where $(x_1,y_1) \in \cQ \triangleq \{(x,y)|x, y \geq 0, x^2+ y^2 \leq 1\}$ 
and $x_2 \in [0, 1]$. Note that this parameterization is different from 
\eqref{eq:wnn} and \eqref{eq:w_lambda} of $\Rnn$. An interpretation stemming 
from \eqref{eq:w1dn}-\eqref{eq:w2dn} is that on the Pareto boundary TX$_2$ 
uses full power, whereas TX$_1$ may not. Inserting 
\eqref{eq:w1dn}--\eqref{eq:w2dn} into \eqref{eq:pq}, we get
\begin{align}
  p_1(\bw_1) &= \left( x_1 \norm{\bPi_{\bh_{12}}\bh_{11}} + y_1 
    \big\Vert\bPi_{\bh_{12}}^\bot \bh_{11}\big\Vert \right)^2 \nonumber \\
  & = (\alpha_1 x_1 + \tilde\alpha_1 y_1)^2, \label{eq:p1dn}\\
  q_2(\bw_1) & =  x_1^2 \frac{|\bh_{12}^H \bh_{11}|^2}
  {\snorm{\bPi_{\bh_{12}}\bh_{11}}}  = g_{12}^2 x_1^2, \label{eq:q2dn} \\
  p_2(\bw_2) & =  x_2^2 \frac{|\bh_{22}^H \bh_{21}|^2}
  {\snorm{\bPi_{\bh_{22}}\bh_{21}}} = g_{22}^2 x_2^2, \label{eq:p2dn}\\
  q_1(\bw_2) & = \left( x_2 \norm{\bPi_{\bh_{22}}\bh_{21}} + \sqrt{1-x_2^2} 
    \big\Vert\bPi_{\bh_{22}}^\bot \bh_{21}\big\Vert \right)^2 \nonumber \\
  & = \left(\beta_2 x_2 + \tilde\beta_2 \sqrt{1-x_2^2}\right)^2, 
  \label{eq:q1dn}
\end{align}
where the parameters $\alpha_1, \alphat_1, \beta_1, \betat_1, g_{12},
g_{22}$ are positive, as defined in Sec.~\ref{sec:systemmodel}. When $x_1$ 
increases, both the power of the desired signal \eqref{eq:p1dn} and the 
interference \eqref{eq:q2dn} increase, whereas $y_1$ only increases the power 
of the desired signal. When $x_2$ increases, the desired signal power 
\eqref{eq:p2dn} increases.

Inserting \eqref{eq:p1dn}--\eqref{eq:q1dn} in 
\eqref{eq:RdnObj}--\eqref{eq:RdnCon1}, replacing the constraint 
\eqref{eq:RdnCon1} in the denominator of the first fraction of 
\eqref{eq:RdnObj}, and taking the square root, we equivalently obtain
\begin{align}
  & \underset{\substack{(x_1,y_1) \in \cQ\\x_2\in[0,1]}}{\text{maximize}}
  && \min \left\{ \frac{\beta_2 x_2+\tilde{\beta}_2 \sqrt{1-x_2^2}}
    {\sqrt{\sigma_1^2(\gamma_1\s + 1)}},
    \frac{g_{22} x_2}  {\sqrt{g_{12}^2 x_1^2 + \sigma_2^2}} \right\} 
  \label{eq:RdnElObj} \\
  &\text{subject to } && \alpha_1 x_1+\tilde{\alpha}_1 y_1 = 
  \sqrt{\gamma_1\s \sigma_1^2}. \label{eq:RdnElCon1}
\end{align}
By Lem. \ref{lem:quasi-concave}, we see that the two fractions in the
objective function \eqref{eq:RdnElObj} are quasi-concave. Since the
minimum of two quasi-concave functions is quasi-concave and
\eqref{eq:RdnElCon1} is linear, it follows that
\eqref{eq:RdnElObj}--\eqref{eq:RdnElCon1} is a quasi-concave problem.

We solve \eqref{eq:RdnElObj}--\eqref{eq:RdnElCon1} in two steps. First we 
solve for $(x_1, y_1)$ and then for $x_2$. We note that $x_1$ appears only in 
the second fraction of \eqref{eq:RdnElObj} and in \eqref{eq:RdnElCon1}, 
whereas $y_1$ appears only in \eqref{eq:RdnElCon1}. The second fraction of 
\eqref{eq:RdnElObj} is monotonously decreasing with $x_1$, for fixed $x_2$, so 
we maximize it by minimizing $x_1$, subject to the constraint 
\eqref{eq:RdnElCon1}, i.e.,
\begin{align}
  &\underset{(x_1,y_1) \in \cQ}{\text{minimize }} &&
  x_1 \label{eq:w1dnObj} \\
  &\text{subject to } &&  y_1 = -\frac{\alpha_1}{\tilde{\alpha}_1} x_1
  + \frac{\sqrt{\gamma_1\s \sigma_1^2}}
  {\tilde{\alpha}_1}. \label{eq:w1dnCon1}  
\end{align}
The solution of this problem can be found by inspection, noting that
the feasible set is the segment of the line \eqref{eq:w1dnCon1} in
$\cQ.$ When $\sqrt{\gamma_1\s \sigma_1^2}/\tilde{\alpha}_1 \leq 1$,
this line segment does not intersect the unit-radius circle; hence,
the optimum value is $x_1\s = 0$. The interpretation of this solution
is that TX$_1$ uses the ZF beamforming vector, but contrary to the
$\Rnn$ case, it may not use full power in order to make interference
cancellation possible. When $\sqrt{\gamma_1\s
  \sigma_1^2}/\tilde{\alpha}_1 > 1$, the line segment intersects the
unit circle in two points; hence, the leftmost is the
optimum. Inserting \eqref{eq:w1dnCon1} into the quadratic equation of
the unit circle, it is straightforward to determine that $x_1\s =
\left(\alpha_1\sqrt{ gamma_1\s\sigma_1^2} - \tilde{\alpha}_1 \sqrt{
    g_{11}^2 - \gamma_1\s \sigma_1^2 } \right)/g_{11}^2$, where we
have used the fact that $g_{11}^2=\alpha_1^2 + \tilde\alpha_1^2$. The
interpretation of this solution is that TX$_1$ uses full power in this
case. 

\begin{figure}[t]
  \begin{center}
    \psfrag{0}[r][r][0.6]{0}
    \psfrag{1}[c][c][0.6]{1}
    \psfrag{2}[r][r][0.6]{2}
    \psfrag{(a)}[c][c][0.6]{(a)}
    \psfrag{(b)}[c][c][0.6]{(b)}
    \psfrag{(c)}[c][c][0.6]{(c)}
    \psfrag{x2}[c][c][0.6]{$x_2$}
    \psfrag{objective value}[c][c][0.6]{Objective value of \eqref{eq:w2dnElObj}}
    \includegraphics[width=\linewidth]{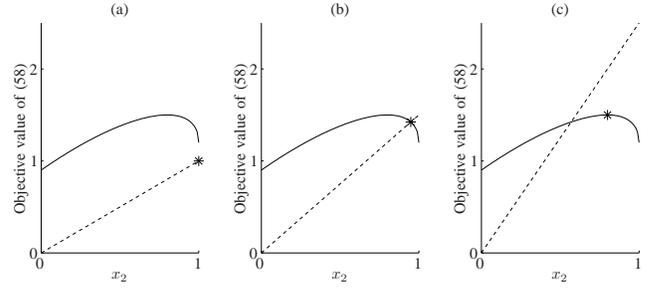}
  \end{center}
  \caption{Illustration of the three different cases in the proof of
    Prop. \ref{prop:Rdn}. The optimal solution is marked with a
    star. (a): $a=1,$ $b=1.2,$ $c=0.9$, (b): $a=1.5,$ $b=1.2,$
    $c=0.9$, (c): $a=2.5,$ $b=1.2,$ $c=0.9$.}
  \label{fig:w2dnOpt}
\end{figure}
Given the optimal $x_1\s$, the optimal $y_1\s$ is determined by 
\eqref{eq:w1dnCon1}, and the problem \eqref{eq:RdnElObj}--\eqref{eq:RdnElCon1} 
only depends on $x_2$, i.e.,
\begin{equation}
  \underset{0 \leq x_2 \leq 1}{\text{maximize }} ~~  \min 
  \left\{\frac{\beta_2 x_2+\tilde{\beta}_2 \sqrt{1-x_2^2}}
    {\sqrt{(\gamma_1\s+1) \sigma_1^2}},
    \frac{g_{22} x_2} {\sqrt{g_{12}^2 (x_1\s)^2 + \sigma_2^2}}
  \right\}. \label{eq:w2dnObj}
\end{equation}
In order to simplify the notation, we define the constants $a
\triangleq g_{22} / \sqrt{(x_1\s)^2 g_{12}^2 +  \sigma_2^2},$ $b
\triangleq \beta_2 / \sqrt{\sigma_1^2(\gamma_1\s + 1)},$ and $c
\triangleq \tilde{\beta}_2 / \sqrt{\sigma_1^2(\gamma_1\s + 1)}.$
Using these constants, we write \eqref{eq:w2dnObj} as
\begin{equation}
  \underset{0 \leq x_2 \leq 1}{\text{maximize }} ~~  \min \left\{b x_2+c 
    \sqrt{1-x_2^2},
    a x_2 \right\}. \label{eq:w2dnElObj}
\end{equation}
Depending on the values of $a,$ $b,$ and $c$, we get three different cases for 
the objective functions in \eqref{eq:w2dnElObj}, as depicted in Fig. 
\ref{fig:w2dnOpt}. In Fig. \ref{fig:w2dnOpt} (a), we have $a\leq b$ and it is 
clear that the optimum is at $x_2\s=1$. The interpretation is that TX$_2$ uses 
the MR beamforming vector. The difference between the cases in Fig. 
\ref{fig:w2dnOpt} (b) and Fig. \ref{fig:w2dnOpt} (c) is whether the 
intersection of the curve with the straight line is to the left or right of 
the maximum of the curve. The curve $b x_2+c \sqrt{1-x_2^2}$ is maximized for 
$x_2 = b/\sqrt{b^2+c^2}.$  The intersection of $b x_2+c \sqrt{1-x_2^2}$ with $a
x_2$ happens for $x_2 = c/\sqrt{c^2+(a-b)^2}.$ In Fig. \ref{fig:w2dnOpt} (b), 
the intersection is to the right of curve's maximum; hence,
\begin{equation}
  \frac{c} {\sqrt{c^2+(a-b)^2}} \geq \frac{b} {\sqrt{b^2+c^2}}
  \Leftrightarrow ab \leq b^2+c^2.
\end{equation}
From Fig. \ref{fig:w2dnOpt} (b), we see that the optimum is at the 
intersection. Hence, we have $x_2\s = c / \sqrt{c^2+(a-b)^2}$. For the case in 
Fig. \ref{fig:w2dnOpt} (c), we see that the optimum is at $x_2\s =
b/\sqrt{b^2+c^2} = \beta/g_{21}$. For the cases depicted in Figs.
\ref{fig:w2dnOpt} (a) and (b), we have $ab \leq b^2+c^2$ and the solution lies
on the line $a x_2$. Therefore, we have $\gamma_2\s = (a x_2\s)^2$. In Fig.
\ref{fig:w2dnOpt} (c), we have $ab > b^2+c^2$ and the optimum lies on the 
curve $b x_2+c \sqrt{1-x_2^2}$. Therefore, we have $\gamma_2\s = b^2+c^2$. The 
interpretation is that TX$_2$ uses a beamforming vector in the direction of 
the crosstalk channel. Note that in Fig. \ref{fig:w2dnOpt}, we depicted the 
scenario of $b > c$. The above analysis does not change if $b \leq c$. We 
summarize the solutions of \eqref{eq:w1dnObj}--\eqref{eq:w1dnCon1} and 
\eqref{eq:w2dnObj} in the following proposition.
\begin{proposition} \label{prop:Rdn}
  The optimal solution $(x_1\s,y_1\s)$ of
  \eqref{eq:w1dnObj}--\eqref{eq:w1dnCon1} is
  \begin{align}
    x_1\s &= \max \left\{0, \frac{\alpha_1}{g_{11}^2}
      \sqrt{\gamma_1\s \sigma_1^2} - \frac{\tilde{\alpha}_1}{g_{11}^2}
      \sqrt{g_{11}^2 - \gamma_1\s \sigma_1^2 }  \right\},
    \label{eq:x1star}\\
    y_1\s &= \left(\sqrt{\gamma_1\s \sigma_1^2}
      -\alpha_1 x_1\s \right) / \tilde{\alpha}_1. \label{eq:y1star}
  \end{align}
  Then, the optimal value of \eqref{eq:RdnObj} is given as 
  \begin{equation}
    \gamma_2\s = \left\{ \begin{array}{ll}
	\frac{\displaystyle g_{22}^2 (x_2\s)^2} {\displaystyle (x_1\s
          g_{12})^2 + \sigma_2^2}, & 
	a \leq b + c^2/b, \\ [1ex] 
        \frac{\displaystyle g_{21}^2} {\displaystyle
          \sigma_1^2(\gamma_1\s+1)}, &  
	a > b + c^2/b
      \end{array} \right. \label{eq:gamma2star}
  \end{equation}
  for
  \begin{align}
    x_2\s &= \left\{ \begin{array}{ll}
        1, & a \leq b, \\ [0.5ex]
        c / \sqrt{c^2+(a-b)^2}, & b < a \leq b + c^2/b, \\[0.5ex]
        \beta_2/g_{21}, & a > b + c^2/b,
      \end{array} \right. \label{eq:x2star}
  \end{align}
  The optimal $(\bw_1\s, \bw_2\s)$ is obtained by inserting
  \eqref{eq:x1star}--\eqref{eq:y1star} and
  \eqref{eq:x2star} into
  \eqref{eq:w1dn}--\eqref{eq:w2dn}.
\end{proposition}
Prop. \ref{prop:Rdn} provides a scheme for finding in closed-form a point on 
the Pareto boundary $\Bdn$, by providing $\gamma_2\s$ as an explicit function 
of $\gamma_1\s$. In Tab. \ref{tab:Rdn_algorithm}, we summarize the proposed 
method for computing the entire $\Bdn.$

Contrary to the $\Rnn$ case, we can obtain the weak, i.e., vertical and 
horizontal, parts of $\Bdn$ by using the method in 
Tab.~\ref{tab:Rdn_algorithm}. The reason is that by using the parameterization 
\eqref{eq:w1dn}--\eqref{eq:w2dn}, we can set $p_i(\bw_i) = 0,$ which is not 
possible for $\Rnn.$ Moreover, it is of interest to analyze the largest value, 
$\gamub_2\dn$, that $\gamma_2\s$ can assume, i.e., when $\gamma_1\s=0$, since 
this brings some insight to $\Rdn$. At this point we have $a = g_{22} / 
\sigma_2,$ $b = \beta_2 / \sigma_1,$ and $c = \tilde{\beta}_2 / \sigma_1.$ 
Therefore, for $\gamma_1\s = 0$ we have,
\begin{equation} \label{eq:gamma2bar_dn}
  \gamub\dn_2 = \left\{ \begin{array}{ll}
      \frac{\displaystyle g_{22}^2} {\displaystyle \sigma_2^2}, &
      \frac{\displaystyle g_{22}} {\displaystyle \sigma_2}
      \leq \frac{\displaystyle \beta_2} {\displaystyle \sigma_1}, \\
      \frac{\displaystyle \tilde{\beta}^2_2 g_{22}^2}
      {\displaystyle \tilde{\beta}^2_2 \sigma_2^2 +(g_{22} \sigma_1 -
	\beta_2 \sigma_2)^2}, &
      \frac{\displaystyle \beta_2} {\displaystyle \sigma_1} <
      \frac{\displaystyle g_{22}} {\displaystyle \sigma_2} \leq
      \frac{\displaystyle g_{21}} {\displaystyle \kappa_2 \sigma_1},
      \\
      \frac{\displaystyle g_{21}^2} {\displaystyle \sigma_1^2}, &
      \frac{\displaystyle g_{22}} {\displaystyle \sigma_2} >
      \frac{\displaystyle g_{21}} {\displaystyle \kappa_2 \sigma_1}.
    \end{array} \right.
\end{equation}
Since $\beta_2 = \kappa_2 g_{21}$, the first case of
\eqref{eq:gamma2bar_dn} corresponds to the scenario where the
crosstalk channel $\bh_{21}$ is strong compared to the direct channel
$\bh_{22}$ and the spatial correlation between $\bh_{21}$ and
$\bh_{22}$ is large. For this scenario, it is optimal for TX$_2$ to
use the MR beamforming vector. The third case of
\eqref{eq:gamma2bar_dn} corresponds to the scenario where $\bh_{21}$
is weak compared to $\bh_{22}$, and the spatial correlation between
$\bh_{21}$ and $\bh_{22}$ is large. For this case, TX$_2$ has to
prioritize the SINR at RX$_1$ and it uses a beamforming vector in the
direction of $\bh_{21}$. The second case of \eqref{eq:gamma2bar_dn} is
somewhere in between the previous extreme cases. For this case, TX$_2$
chooses its beamforming vector such that both the RXs get the
same SINR. To conclude, the highest rate link 2 can achieve in $\Rdn$ is
$\Rub\dn_2 = \log_2(1+\gamub\dn_2)$.

\begin{table}[ht]
  \begin{center}
    \begin{tabular}{|rl|}
      \hline
      1: & \textbf{Input and output:} same as in Tab. \ref{tab:Rnn_num} 
      \\
      2: & $\br_1 = \left[0: \Rub\dn_1/(M-1): \Rub\dn_1\right]$ \\
      3: & \textbf{for } $k = 1:M$ \\
      4: & ~~~ $\gamma_1\s = 2^{\br_1(k)}-1$ \\
      5: & ~~~ Compute $x_1\s$ and $y_1\s$ using \eqref{eq:x1star} and 
      \eqref{eq:y1star} \\
      6: & ~~~ Compute $x_2\s$ using \eqref{eq:x2star} \\
      7: & ~~~ Compute $\bw_1\s$ and $\bw_2\s$ using \eqref{eq:w1dn} and 
      \eqref{eq:w2dn} \\
      8: & ~~~ Compute $\gamma_2\s$ using \eqref{eq:gamma2star} \\
      9: & ~~~ $\br_2(k) = R_2\s = \log_2(1+\gamma_2\s)$ \\
      10: & \textbf{end} \\
      \hline
    \end{tabular}
  \end{center}
  \caption{Closed-form method to compute $\Bdn$}
  \label{tab:Rdn_algorithm}
\end{table}

\section{Both RXs Decode the Interference}
\label{sec:Rdd}

In this section, we propose a computationally efficient numerical method to 
compute $\Bdd$ which is similar in logic to the one given in Sec. 
\ref{sec:WiOpt} for $\Bnn$ and also utilizes intermediate results from the 
method in Sec. \ref{sec:Rdn} for $\Bdn$. The method proposed herein improves 
the corresponding one of \cite{Lindblom-2011-CAMSAP} in two ways: a) the 
number of variables is decreased from four real variables to a single one and 
b) a single instance of two quasi-concave subproblems needs to be solved 
instead of a sequence of convex feasibility problems.

Inserting the rate expressions \eqref{eq:R1dd} and \eqref{eq:R2dd} in the 
optimization problem \eqref{eq:RoptObj}--\eqref{eq:RoptCon} and equivalently 
reformulating the rate maximization to SINR maximization problem, we obtain
\begin{align}
  &\underset{ (\bw_1,\bw_2) \in \cW^2}{\text{maximize }} &&
  \min\left\{ \frac{\displaystyle p_2(\bw_2)} {\displaystyle
      \sigma_2^2}, \frac{\displaystyle q_1(\bw_2)} {\displaystyle
      p_1(\bw_1) + \sigma_1^2} \right\} \label{eq:RddObj} \\
  &\text{subject to } && p_1(\bw_1) / \sigma_1^2 \geq \gamma_1\s, 
  \label{eq:RddCon1} \\
  & && \frac{q_2(\bw_1)}  {p_2(\bw_2) + \sigma_2^2} \geq \gamma_1\s, 
  \label{eq:RddCon2}
\end{align}
where \eqref{eq:RddCon1}--\eqref{eq:RddCon2} follow from the epigraph 
formulation of \eqref{eq:R1dd}, see \cite[Ch. 3]{Boyd-ConvexOptimization}. The 
formulation \eqref{eq:RddObj}--\eqref{eq:RddCon2} is nonconvex, since the 
constraints are fractions of quadratic functions.

In \cite{Ho-2010-Asilomar}, it was shown that the PO beamforming vectors of 
the  $\Rdd$ region can be parameterized as
\begin{align}
  \bw_i(x_i,y_i) & = x_i \frac{\bPi_{\bh_{ii}} 
    \bh_{ij}}{\norm{\bPi_{\bh_{ii}} \bh_{ij}}} + y_i 
  \frac{\bPi_{\bh_{ii}}^\bot \bh_{ij}} {\big\Vert\bPi_{\bh_{ii}}^\bot 
    \bh_{ij}\big\Vert} \label{eq:widd}
\end{align}
for $i,j=1,2$ and $j\neq i$, where $(x_i,y_i) \in \cQ $. Note that this 
parameterization is different from \eqref{eq:wnn} and \eqref{eq:w_lambda} of 
$\Rnn$ and \eqref{eq:w1dn}--\eqref{eq:w2dn} of $\Rdn$. Inserting 
\eqref{eq:widd} into \eqref{eq:pq}, we get
\begin{align}
  p_i(\bw_i) &=  x_i^2 \frac{|\bh_{ii}^H \bh_{ij}|^2}
  {\snorm{\bPi_{\bh_{ii}}\bh_{ij}}} = g_{ii}^2 x_i^2, 
  \label{eq:pidd}\\
  q_i(\bw_j) &= \left( x_j \norm{\bPi_{\bh_{jj}}\bh_{ji}} + y_j 
    \big\Vert\bPi_{\bh_{jj}}^\bot \bh_{ji}\big\Vert \right)^2 \nonumber \\
  & = \left(\beta_j x_j + \betat_j y_j\right)^2,.
  \label{eq:qidd}
\end{align}
where the parameters $g_{ii},\beta_i,\betat_j$ are defined in Sec. 
\ref{sec:systemmodel}. From \eqref{eq:widd}, we see that the PO beamforming 
vectors of both TXs do not necessarily use all available power. However, 
without loss of optimality, we can assume that full power is used at optimum.
This is so because increasing $y_i$ increases the interference \eqref{eq:qidd} 
but does not affect the desired signal power \eqref{eq:pidd}.
The effect of increasing $y_i$, beyond the optimal solution $y_i\s$, is to 
only make the constraint \eqref{eq:RddCon2} looser at optimum, without 
decreasing the objective value \eqref{eq:RddObj}. The interpretation is that 
we can increase the interference arbitrarily, since it will be canceled by the 
RXs. Hence, we can increase it until the power constraint is met with 
equality, i.e., set $y_i = \sqrt{1-x_i^2}$.

Inserting \eqref{eq:pidd}--\eqref{eq:qidd} in 
\eqref{eq:RddObj}--\eqref{eq:RddCon2}, taking the square root of
objective and constraints, and introducing the nonnegative auxiliary
variable $z$, we equivalently obtain
\begin{align}
  &\underset{0 \leq x_1,x_2 \leq 1, \enspace z \geq 0}
  {\text{maximize}} && z \label{eq:RddElObj} \\
  &\text{subject to} && g_{22} x_2 / \sigma_2 \geq z,\label{eq:RddElCon1} \\
  & && \frac{ \beta_2 x_2 + \betat_2 \sqrt{1-x_2^2}} {\sqrt{g_{11}^2 x_1^2 + 
      \sigma_1^2}} \geq z, \label{eq:RddElCon2} \\
  & && g_{11} x_1 / \sigma_1 \geq \sqrt{\gamma_1\s}, \label{eq:RddElCon3} \\
  & && \frac{\beta_1 x_1 + \betat_1 \sqrt{1-x_1^2}} {\sqrt{g_{22}^2 x_2^2 + 
      \sigma_2^2}} \geq \sqrt{\gamma_1\s}. \label{eq:RddElCon4}
\end{align}
In order to further simplify the notation, we define
\begin{align}
  \ut_i(x_i) & \triangleq \beta_i x_i + \betat_i \sqrt{1-x_i^2},
  \label{eq:ut_def} \\
  \vt_i(x_i) & \triangleq \sqrt{g_{ii}^2 x_i^2 + \sigma_i^2}. 
  \label{eq:vt_def}
\end{align}
Problem \eqref{eq:RddElObj}--\eqref{eq:RddElCon4} is feasible for $\gamma_1\s 
\in [0, \gamub\dd_1]$. The rate $\Rub\dd_1 = \log_2(1+\gamub\dd_1)$ is the 
highest rate of link TX$_1\rightarrow$RX$_1$ that can be decoded by both RXs, 
achieved when TX$_2$ does not transmit.
We determine $\gamub\dd_1$ as
\begin{equation} \label{eq:gamma1bar_dd}
  \gamub\dd_1 =  \left( \underset{0 \leq x_1 \leq 1}{\text{maximize}} 
    \enspace 
    \min \left\{ \frac{g_{11} x_1} {\sigma_1}, \frac{\beta_1 x_1 + \betat_1 
        \sqrt{1-x_1^2}} {\sigma_2} \right\} \right)^2.
\end{equation}
The maximization in \eqref{eq:gamma1bar_dd} is similar to 
\eqref{eq:w2dnElObj}, so we can solve it using \eqref{eq:gamma2bar_dn}. By 
interchanging the indices in
\eqref{eq:gamma1bar_dd}, we can find $\gamub\dd_2$.

Due to the variable $z,$ \eqref{eq:RddElCon2} does not define a convex set. 
Hence, \eqref{eq:RddElObj}--\eqref{eq:RddElCon4} is neither a concave nor a 
quasi-concave problem as it stands. But, by using the epigraph formulation, 
\eqref{eq:RddElObj}--\eqref{eq:RddElCon4} can be equivalently reformulated 
into a quasi-concave problem. By studying the KKT conditions of 
\eqref{eq:RddElObj}--\eqref{eq:RddElCon4}, we identify two cases and apply to 
each of them techniques introduced in Secs. \ref{sec:WiOpt} and \ref{sec:Rdn}, 
respectively. The fact that the gradient of the Lagrange function of
\eqref{eq:RddElObj}--\eqref{eq:RddElCon4} vanishes at the optimum,
gives the KKT conditions
\begin{align}
  &\frac{\partial \cL} {\partial z} = 1 - \mu_1 - \mu_2 = 0, 
  \label{eq:Rdd_KKT_t}\\
  &\frac{\partial \cL} {\partial x_1} =
  -\mu_2  \frac{ \vt_1'(x_1) \ut_2(x_2)} {\vt_1^2(x_1)} + \mu_3 
  \frac{g_{11}}{\sigma_1} +
  \mu_4 \frac{\ut_1'(x_1)} {\vt_2(x_2)} = 0, \label{eq:Rdd_KKT_x1} \\
  &\frac{\partial \cL} {\partial x_2} = \mu_1 \frac{g_{22}}{\sigma_2} +
  \mu_2 \frac{\ut_2'(x_2)} {\vt_1(x_1)} - \mu_4 \frac{ \vt_2'(x_2)
    \ut_2(x_2)} {\vt_2^2(x_2)} = 0, \label{eq:Rdd_KKT_x2}
\end{align}
where $\mu_i \geq 0,$ $i\in\{1,2,3,4\}$ are the Lagrange multipliers
of constraints \eqref{eq:RddElCon1}--\eqref{eq:RddElCon4},
respectively. First, we observe that we can have
$\mu_3 = \mu_4 = 0$ only when $x_1 = 0$. This is the case only when
$\gamma_1\s = 0.$ Hence, for every other point we have either
$\mu_3 > 0$ or $\mu_4 > 0,$ corresponding to equality in
\eqref{eq:RddElCon3} and \eqref{eq:RddElCon4}, respectively. Next, for
each case, we change the corresponding inequality in
\eqref{eq:RddElObj}--\eqref{eq:RddElCon4} to equality. We solve the
two programs separately and compare the solutions. The solution with
the highest optimal value will yield the optimum of
\eqref{eq:RddElObj}--\eqref{eq:RddElCon4}.

For the case of equality in \eqref{eq:RddElCon3}, it immediately
follows that $x_1\s = \sqrt{\gamma_1\s \sigma_1^2 / g_{11}^2}$. From
\eqref{eq:RddElCon4}, we see that the problem is feasible only if
$\beta_1 x_1\s + \betat_1 \sqrt{1-(x_1\s)^2} \geq \sqrt{\gamma_1\s 
\sigma_2^2}$. Note that this is always the case if $\gamma_1\s \leq 
\gamub\dd_1$. Given that \eqref{eq:RddElObj}--\eqref{eq:RddElCon4} is feasible 
with equality in \eqref{eq:RddElCon3}, we have the problem
\begin{align}
  &\underset{0 \leq x_2 \leq 1} {\text{maximize}}
  & & \min\left\{ \frac{ g_{22} x_2} { \sigma_2}, \frac{ \ut_2(x_2)} { 
      \sqrt{\gamma_1\s + \sigma_1^2}} \right\} \label{eq:RddEl_Prob1} \\
  &\text{subject to} & & x_2 \leq \frac{1} {g_{22} \sqrt{\gamma_1\s}} \sqrt{
    \ut_1^2(x_1\s)-\gamma_1\s\sigma_2^2}. \label{eq:RddElCon_Prob1}
\end{align}
Note that $x_2$ is real and non-negative whenever the structure of 
\eqref{eq:RddEl_Prob1}--\eqref{eq:RddElCon_Prob1} is similar to 
\eqref{eq:w2dnObj}. The only difference is the extra constraint 
\eqref{eq:RddElCon_Prob1} which yields a tighter upper bound for $x_2$. Hence, 
we can use Prop. \ref{prop:Rdn} to find $x_2\s$ by using coefficients 
$\tilde{a} \triangleq g_{22}/\sigma_2,$ $\tilde{b} \triangleq \beta_2 / 
\sqrt{\gamma_1\s + \sigma_1^2},$ and $\tilde{c} \triangleq \betat_2 / 
\sqrt{\gamma_1\s + \sigma_1^2},$ in place of $a,$ $b,$ and $c$, respectively.

For the case of equality in \eqref{eq:RddElCon4}, we get
\begin{equation} \label{eq:Rdd_x2}
x_2 = \frac{1} {g_{22} \sqrt{\gamma_1\s}}
\sqrt{\ut_1^2(x_1)-\gamma_1\s\sigma_2^2} \triangleq
\tilde{w}(x_1).
\end{equation}
Inserting \eqref{eq:Rdd_x2} in \eqref{eq:RddElObj}--\eqref{eq:RddElCon4} 
yields the problem
\begin{align}
  &\underset{0 \leq x_1 \leq 1} {\text{maximize}} & &
  \min\left\{ s_1(x_1), s_2(x_1) \right\}
  \label{eq:RddEl_Prob2_Obj} \\
  &\text{subject to} & & g_{11} x_1 / \sigma_1 \geq
  \sqrt{\gamma_1\s}, \label{eq:RddEl_Prob2_Con1} \\
  & & & \ut_1(x_1) \geq \sqrt{\gamma_1\s \sigma_2^2}, 
  \label{eq:RddEl_Prob2_Con3} \\
  & & & \ut_1(x_1) \leq \sqrt{\gamma_1\s(g_{22}^2+\sigma_2^2)},  
  \label{eq:RddEl_Prob2_Con2}   
\end{align}
where
\begin{align}
  s_1(x_1) & \triangleq g_{22} \tilde{w}(x_1) /
  \sigma_2, \label{eq:Rdd_s1} \\ 
  s_2(x_1) & \triangleq \ut_2 \left( \tilde{w}(x_1) \right) / \vt_1(x_1).
  \label{eq:Rdd_s2}
\end{align}
The constraints \eqref{eq:RddEl_Prob2_Con3} and \eqref{eq:RddEl_Prob2_Con2} 
correspond to $x_2 \geq 0$ and $x_2 \leq 1,$ respectively. Constraint 
\eqref{eq:RddEl_Prob2_Con1} is satisfied if $g_{11}/\sigma_1^2 \geq 
\gamma_1\s$, constraint \eqref{eq:RddEl_Prob2_Con3} is satisfied if 
$\ut_1(\kappa_1) = g_{12}^2 \geq \sqrt{\gamma_1\s \sigma_2^2} $ and constraint 
\eqref{eq:RddEl_Prob2_Con2} is satisfied if $\ut_1(0) = \betat_1 \leq 
\sqrt{\gamma_1\s(g_{22}^2+\sigma_2^2)}$. Hence, 
\eqref{eq:RddEl_Prob2_Obj}--\eqref{eq:RddEl_Prob2_Con2} is feasible when 
\begin{equation} \label{eq:Rdd_Prob2Feasible}
  \frac{(1-\kappa_1^2) g_{12}^2} {g_{22}^2+\sigma_2^2} \leq \gamma_1\s 
  \leq \min \left\{ \frac{g_{11}^2}{\sigma_1^2}, 
    \frac{g_{12}^2}{\sigma_2^2}\right\}.
\end{equation}
By comparing the RHS of \eqref{eq:Rdd_Prob2Feasible} with 
\eqref{eq:gamma1bar_dd}, we see that $\gamub\dd_1 \leq \min\{g_{11}^2/ 
\sigma_1^2, g_{12}^2/ \sigma_2^2\}$. While 
\eqref{eq:RddEl_Prob1}--\eqref{eq:RddElCon_Prob1} is feasible for all 
$\gamma_1\s \leq \gamub\dd_1$, the optimization 
\eqref{eq:RddEl_Prob2_Obj}--\eqref{eq:RddEl_Prob2_Con2} is feasible only for a 
smaller subset as already illustrated by \eqref{eq:Rdd_Prob2Feasible}. The 
optimization \eqref{eq:RddEl_Prob2_Obj}--\eqref{eq:RddEl_Prob2_Con2} will be 
solved using a method similar to that used for solving 
\eqref{eq:Rnn_scalar_opt} for $\Bdd$ in Sec.~\ref{sec:WiOpt}. First, we show 
that we do not need to solve 
\eqref{eq:RddEl_Prob2_Obj}--\eqref{eq:RddEl_Prob2_Con2} for all $\gamma_1\s$ 
satisfying \eqref{eq:Rdd_Prob2Feasible}. By revisiting the KKT conditions 
\eqref{eq:Rdd_KKT_t}--\eqref{eq:Rdd_KKT_x2}, we see that if $\mu_4 > 0$ and 
$x_1 > \kappa_1$, then we must have $\mu_3 > 0$ as well. This follows since 
$\ut_1'(x_1) < 0$ for $x_1 > \kappa_1$. But the case of $\mu_3 > 0$ is already 
covered by \eqref{eq:RddEl_Prob1}--\eqref{eq:RddElCon_Prob1}. Hence, if 
$\gamma_1\s > g_{11}^2 \kappa_1^2/\sigma_1^2,$ it suffices to solve 
\eqref{eq:RddEl_Prob1}--\eqref{eq:RddElCon_Prob1} and we set the optimal value 
of \eqref{eq:RddEl_Prob2_Obj}--\eqref{eq:RddEl_Prob2_Con2} to zero. Therefore, 
in the following, we only consider the upper bound $\gamma_1\s \leq g_{11}^2 
\kappa_1^2/\sigma_1^2 $ and lower bound of \eqref{eq:Rdd_Prob2Feasible}. 
Second, we determine upper and lower bounds on $x_1.$ Since $\ut_1(x_1)$ is 
non-decreasing for $x_1 \leq \kappa_1$, the constraint yields an upper bound 
on the optimal $x_1$; namely $x_1 \leq \xub_1$, where 
\begin{equation} \label{eq:Rdd_xub}
  \hspace*{-2mm} \xub_1 \triangleq \left\{ \begin{array} {ll}
      \kappa_1 \sqrt{ \displaystyle \frac{\gamma_1\s} {\tilde{\gamma}_1\MR}} 
      -\sqrt{1-\kappa_1^2} 
      \sqrt{ \displaystyle 1 - \frac{\gamma_1\s} {\tilde{\gamma}_1\MR}}, & \!\!
      \gamma_1\s \leq \tilde{\gamma}_1\MR, \\
      \kappa_1, & \!\! \gamma_1\s > \tilde{\gamma}_1\MR,
    \end{array} \right. \!\!\!\!
\end{equation}
where $\tilde{\gamma}\MR_1 \triangleq g_{12}^2/(g_{22}^2+\sigma_2^2)$ is the 
SINR at RX$_2$ when it decodes the interference while TX$_1$ and TX$_2$ 
transmit in the MR directions of $\bh_{12}$ and $\bh_{22},$ respectively. The 
constraints \eqref{eq:RddEl_Prob2_Con1} and \eqref{eq:RddEl_Prob2_Con3} yield 
a lower bound $x_1 \geq \xlb_1$, where 
\begin{equation} \label{eq:Rdd_xlb}
  \xlb_1 \triangleq \max \left\{\frac{\sqrt{\gamma_1\s \sigma_1^2}}{g_{11}}, 
    \kappa_1 \sqrt{ \frac{\sigma_2^2 \gamma_1\s}{g_{12}^2}} - 
    \sqrt{1-\kappa_1^2} \sqrt{ 1- \frac{\sigma_2^2 \gamma_1\s} 
      {g_{12}^2}}\right\}.
\end{equation}

Next, we show that the objective function \eqref{eq:RddEl_Prob2_Obj} is 
quasi-concave for $x_1 \in [\xlb_1,\xub_1]$. We note that the minimum 
of two quasi-concave functions is quasi-concave \cite[Ch. 
3.4]{Boyd-ConvexOptimization}. The function \eqref{eq:Rdd_s1} is on the same 
form as \eqref{eq:Rnn_x2} and hence, it is concave. So, if \eqref{eq:Rdd_s2} 
is quasi-concave, then \eqref{eq:RddEl_Prob2_Obj} is quasi-concave. We note 
that \eqref{eq:Rdd_s2} has the same structure as the objective function of 
\eqref{eq:Rnn_scalar_opt}. Hence it follows from Lem. \ref{lem:stationary} 
that \eqref{eq:Rdd_s2} is quasi-concave.

Since \eqref{eq:RddEl_Prob2_Obj} is quasi-concave, we can use a gradient 
method similar to the respective one for $\Bnn$, presented in Tab. 
\ref{tab:Rnn_num}. The proposed method to compute $\Bdd$ is sketched in Tab. 
\ref{tab:Rdd_num} and differs in the following points to the one for $\Bnn$. 
First, for $\Bdd$ we have to solve two optimization problems, which we do 
separately. We denote $x_{11}\s,$ $x_{21}\s$ and $\gamma_{21}\s$ the optimal 
solution and value, respectively, of 
\eqref{eq:RddEl_Prob1}--\eqref{eq:RddElCon_Prob1}, and $x_{12}\s,$ $x_{22}\s$ 
and $\gamma_{22}\s$ the optimal solution and value, respectively, of 
\eqref{eq:RddEl_Prob2_Obj}--\eqref{eq:RddEl_Prob2_Con2}. The solution to the 
subproblem that yields the highest optimal value is declared the solution to 
\eqref{eq:RddElObj}--\eqref{eq:RddElCon4}. If both subproblems are infeasible, 
we set the optimal value to zero.  Second, we maximize the minimum of two 
functions. Therefore, in lines 13--17, we let the gradient of the objective 
function \eqref{eq:RddEl_Prob2_Obj}, denoted by $\Delta$, at a point $x_1$, 
take as value the minimum of the derivatives of functions $s_1(x_1)$ and 
$s_2(x_2)$. In lines 21--22, we use $\st(x_1) \triangleq \min\left\{ s_1(x_1), 
s_2(x_1) \right\}.$ Except for these points, the method works as for $\Bnn$.

\begin{table}[ht]
  \begin{center}
    \begin{tabular}{|rl|}
      \hline
      1: & \textbf{Input and output:} same as in Tab. \ref{tab:Rnn_num} 
      \\
      2: & \textbf{Output: } $\Bdd$ given by vectors $\br_1,
      \br_2 \in \bbR^M$ \\
      3: & $\br_1 = [0: \Rub\dd/(M-1): \Rub\dd_1]$ \\
      4: & $\br_2(1) = \Rub\dd_2,$ $x_1\s = 0$ \\
      5: & \textbf{for } $k = 2:M$ \\
      6: & ~~~ $\gamma_1\s = 2^{\br_1(k)}-1$ \\
      7: & ~~~ Solve \eqref{eq:RddEl_Prob1}--\eqref{eq:RddElCon_Prob1}
      using Prop. \ref{prop:Rdn} $\Rightarrow x_{11}\s, x_{21}\s,
      \gamma_{21}\s$ \\
      8: & ~~~ \textbf{if} $(1-\kappa_1)^2 g_{12}^2 
      /(g_{22}^2+\sigma_22) 
      \leq \gamma_1\s \leq g_{11}^2 \kappa_1^2 / 
      \sigma_1^2$ \\
      9: & ~~~~~~ Compute $\xub_1$ and $\xlb_1$ using \eqref{eq:Rdd_xub} 
      and 
      \eqref{eq:Rdd_xlb} \\
      10: & ~~~~~~ $x_{12}^{(0)} = [x_1\s]_{\xlb_1}^{\xub_1}$ \\
      11: & ~~~~~~ $l = 0$ \\
      12: & ~~~~~~ \textbf{repeat } \\
      13: & ~~~~~~~~~ \textbf{if} $s_1(x_{12}^{(l)}) \leq 
      s_2(x_{12}^{(l)})$  \\
      14: & ~~~~~~~~~~~~ $\Delta = s_1'(x_{12}^{(l)})$ \\
      15: & ~~~~~~~~~ \textbf{else} \\
      16: & ~~~~~~~~~~~~ $\Delta = s_2'(x_{12}^{(l)})$ \\
      17: & ~~~~~~~~~ \textbf{end} \\
      18: & ~~~~~~~~~ Determine step size $t$ \\
      19: & ~~~~~~~~~ $x_{12}^{(l+1)} = [x_{12}^{(l)}+t 
      \Delta]_{\xlb_1}^{\xub_1}$\\
      20: & ~~~~~~~~~ $l \leftarrow l+1$ \\
      21: & ~~~~~~ \textbf{until } 
      $\big|\st(x_{12}^{(l)})-\st(x_{12}^{(l-1)})\big| < \epsilon$ \\
      22: & ~~~~~~ $\gamma_{22}\s = \st^2 (x_{12}^{(l)})$  \\
      23: & ~~~~~~ Compute $x_{22}$ using \eqref{eq:Rdd_x2} \\
      24: & ~~~ \textbf{else} \\
      25: & ~~~~~~ $\gamma_{22}\s = 0$ \\
      26: & ~~~ \textbf{end} \\
      27: & ~~~ \textbf{if} $\gamma_{21}\s \geq \gamma_{22}\s$ \\
      28: & ~~~~~~ $\gamma_2\s = \gamma_{21}\s$, $x_1\s = x_{11}\s$, 
      $x_2\s = x_{21}\s$ \\
      29: & ~~~ \textbf{else} \\ 
      30: & ~~~~~~ $\gamma_2\s = \gamma_{22}\s$, $x_1\s = x_{12}\s$, 
      $x_2\s = x_{22}\s$ \\
      31: & ~~~ \textbf{end} \\
      32: & ~~~ Compute $\bw_i\s$ using \eqref{eq:widd} \\
      33: & ~~~ Compute $\br_2(k) = R_2\s = \log_2 \left(1 + \gamma_2\s 
      \right)$ \\
      34: & \textbf{end} \\
      \hline
    \end{tabular}
  \end{center}
  \caption{Numerical method to compute $\Bdd$}
  \label{tab:Rdd_num}
\end{table}

\section{Numerical Illustrations}
\label{sec:numerical}

Here we illustrate how the channel parameters $g_{ij},$ $\kappa_i,$ $i,j=1,2$ 
affect the shape of the rate regions. By choosing these parameters in a 
controlled way, instead of randomly drawing channel vectors, we can illustrate 
interesting properties of the four rate regions. Also, we provide an analysis 
of the computational complexity of the proposed methods.

\begin{figure}[H]
  \begin{center}
    \psfrag{0}[r][r][0.8]{0}
    \psfrag{1}[c][c][0.8]{1}
    \psfrag{2}[r][r][0.8]{2}
    \psfrag{3}[r][r][0.8]{3}
    \psfrag{4}[c][c][0.8]{4}
    \psfrag{5}[r][r][0.8]{5}
    \psfrag{R1 [bpcu]}[c][c][0.8]{$R_1$ [bpcu]}
    \psfrag{R2 [bpcu]}[c][c][0.8]{$R_2$ [bpcu]}
    \psfrag{Rnn}[c][c][0.8]{$\Rnn$}
    \psfrag{Rnd}[c][c][0.8]{$\Rnd$}
    \psfrag{Rdn}[c][c][0.8]{$\Rdn$}
    \psfrag{Rdd}[c][c][0.8]{$\Rdd$}
    \psfrag{g11=1, g12=2, g21=2, g22=1, K1=0.3, K2=0.3}[c][c][0.1]{}
    \includegraphics[width=\linewidth]{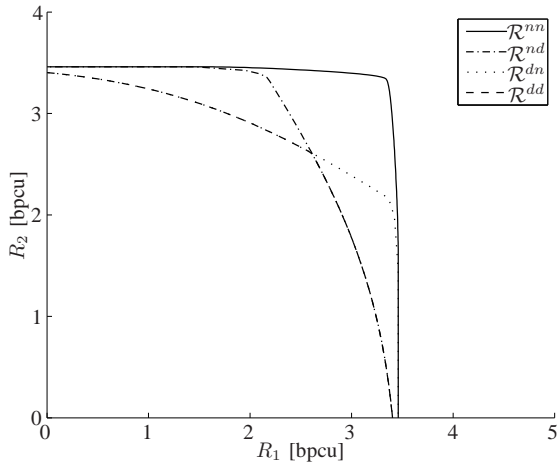}
  \end{center}
  \caption{Rate regions for $g_{11}=g_{22}=1$, $g_{12}=g_{21}=2$, 
    $\kappa_1=\kappa_2=0.3$}
  \label{fig:sym_LowCorr}
\end{figure}

\begin{figure}[H]
  \begin{center}
    \psfrag{0}[r][r][0.8]{0}
    \psfrag{1}[c][c][0.8]{1}
    \psfrag{2}[r][r][0.8]{2}
    \psfrag{3}[r][r][0.8]{3}
    \psfrag{4}[c][c][0.8]{4}
    \psfrag{5}[r][r][0.8]{5}
    \psfrag{R1 [bpcu]}[c][c][0.8]{$R_1$ [bpcu]}
    \psfrag{R2 [bpcu]}[c][c][0.8]{$R_2$ [bpcu]}
    \psfrag{Rnn}[c][c][0.8]{$\Rnn$}
    \psfrag{Rnd}[c][c][0.8]{$\Rnd$}
    \psfrag{Rdn}[c][c][0.8]{$\Rdn$}
    \psfrag{Rdd}[c][c][0.8]{$\Rdd$}
    \psfrag{R1sR2s}[l][l][0.8]{$(R_1\s,R_2\s)$}
    \psfrag{RubRlb}[l][l][0.8]{$(\Rub_1^{nn},\Rlb_2^{nn})$}
    \psfrag{RlbRub}[l][l][0.8]{$(\Rlb_1^{nn},\Rub_2^{nn})$}
    \psfrag{g11=1, g12=2, g21=2, g22=1, K1=0.85, K2=0.85}[c][c][0.1]{}
    \includegraphics[width=\linewidth]{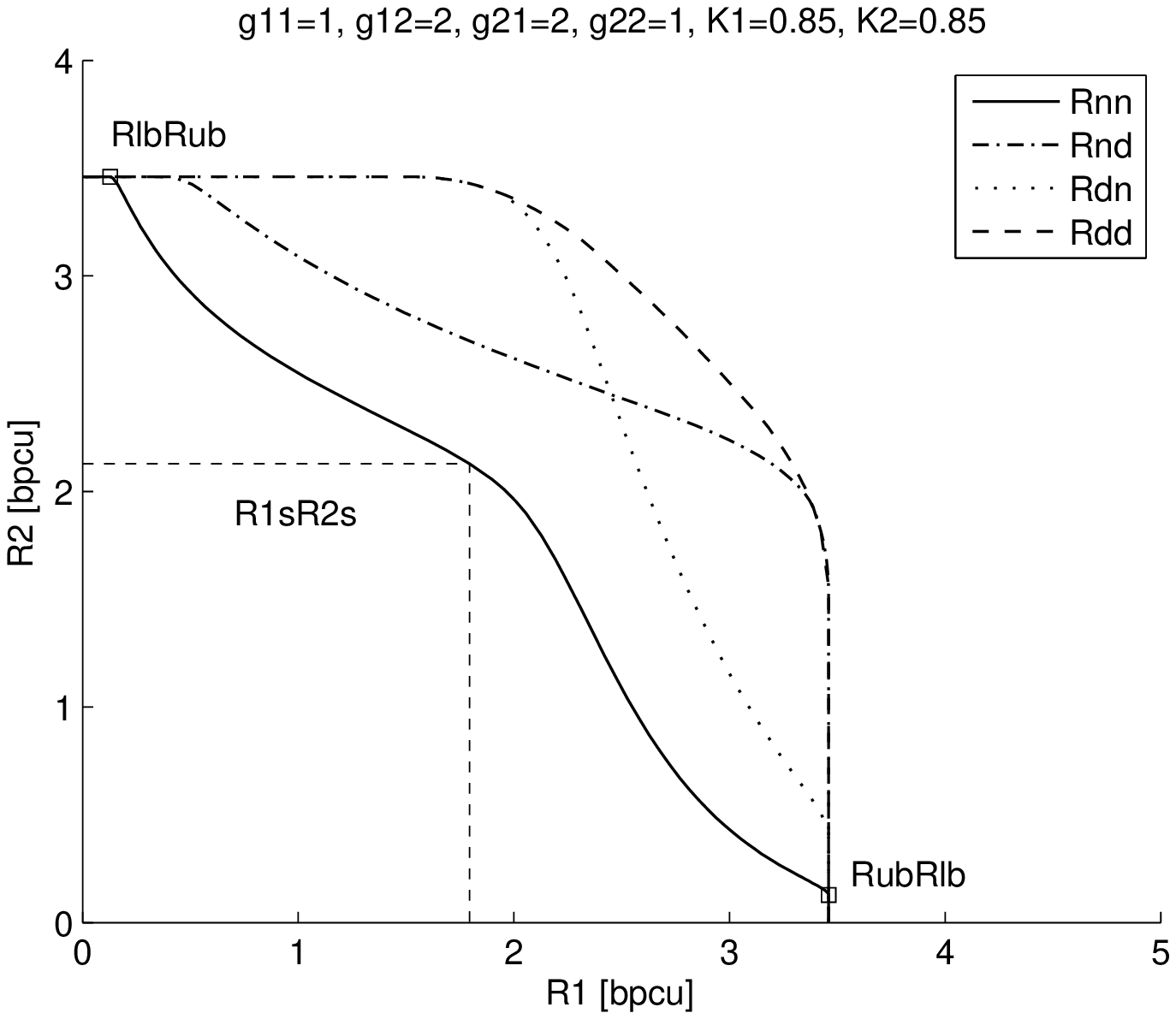}
  \end{center}
  \caption{Rate regions for $g_{11}=g_{22}=1$, $g_{12}=g_{21}=2$, 
    $\kappa_1=\kappa_2=0.85$}
  \label{fig:sym_HighCorr}
\end{figure}

\begin{figure}[H]
  \begin{center}
    \psfrag{0}[r][r][0.8]{0}
    \psfrag{1}[c][c][0.8]{1}
    \psfrag{2}[r][r][0.8]{2}
    \psfrag{3}[r][r][0.8]{3}
    \psfrag{4}[c][c][0.8]{4}
    \psfrag{5}[r][r][0.8]{5}
    \psfrag{R1 [bpcu]}[c][c][0.8]{$R_1$ [bpcu]}
    \psfrag{R2 [bpcu]}[c][c][0.8]{$R_2$ [bpcu]}
    \psfrag{Rnn}[c][c][0.8]{$\Rnn$}
    \psfrag{Rnd}[c][c][0.8]{$\Rnd$}
    \psfrag{Rdn}[c][c][0.8]{$\Rdn$}
    \psfrag{Rdd}[c][c][0.8]{$\Rdd$}
    \psfrag{g11=1, g12=2, g21=2, g22=1, K1=0.85, K2=0.3}[c][c][0.1]{}
    \includegraphics[width=\linewidth]{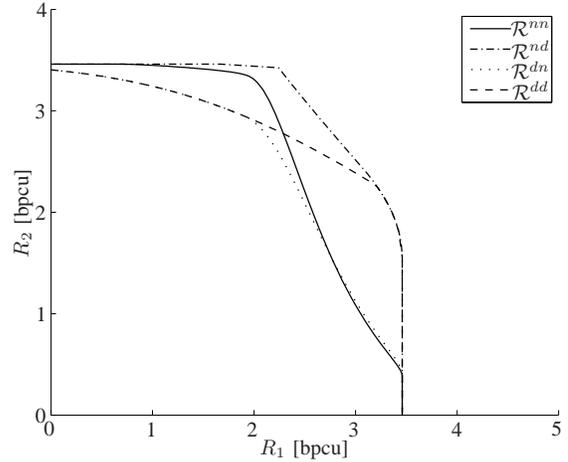}
  \end{center}
  \caption{Rate regions for $g_{11}=g_{22}=1$, $g_{12}=g_{21}=2$, 
    $\kappa_1=0.85$, $\kappa_2=0.3$}
  \label{fig:sym_DiffCorr}
\end{figure}

In Figs. \ref{fig:sym_LowCorr}--\ref{fig:sym_DiffCorr}, we illustrate the 
scenario where the channel gains are symmetric with $g_{11} = g_{22} = 1$ and 
$g_{12} = g_{21} = 2$. That is, the crosstalk channels are stronger than the 
direct channels. In Fig. \ref{fig:sym_LowCorr}, we have $\kappa_1 = \kappa_2 = 
0.3$, which corresponds to a low spatial correlation of amongst the direct and 
crosstalk channels. We see that, even though the crosstalk channel gains are 
high, $\Rnn$ is almost rectangular and all the other regions are contained in 
$\Rnn$. In this case, there is no need of cancel out interference; it costs 
too much in terms of useful signal power to create extra interference in order 
to enable interference cancellation.

We illustrate the other extreme case in Fig. \ref{fig:sym_HighCorr}. Here we 
have $\kappa_1=\kappa_2=0.85$, which corresponds to the case where the angle 
between the direct and crosstalk channel vectors is small. We see that all the 
other regions are contained in $\Rdd$. The combination of strong crosstalk 
channels and high spatial correlation, entails that the cost of boosting 
interference in order enable interference cancellation is very small.

In Fig. \ref{fig:sym_DiffCorr}, we depict the case of $\kappa_1=0.85$ and 
$\kappa_2=0.3$, i.e., the channels from TX$_1$ and TX$_2$ have high and low 
spatial correlation, respectively. In this case we have $\cR = \Rnd$. The 
reason is that RX$_2$ experiences high interference and has no problem to 
decode it. On the other hand RX$_1$ experiences low interference, so it is 
better to treat it as noise.

For both Fig. \ref{fig:sym_LowCorr} and Fig. \ref{fig:sym_DiffCorr}, we see 
that $\Rdd \subseteq \Rdn \bigcup \Rnd$. This is something that we frequently 
observe when the channels are i.i.d. Rayleigh and SNR is around 0 dB. The 
explanation is that for $\Rdd$ both links have to sacrifice part of the 
desired signal power in order to enable interference cancellation.


\subsection{Computational Complexity}
\label{sec:complexity}
Here, we consider the computational complexity of the proposed methods. We 
give both order expressions and the number of flops required to find the 
boundaries for the regions in 
Figs.~\ref{fig:sym_LowCorr}--\ref{fig:sym_DiffCorr}. In 
Tab.~\ref{tab:Complexity}, we compare the complexity of the proposed methods 
with that of the brute-force methods. 

The complexity of the proposed methods depends on the number of grid points, 
$M$, of each parameter. For the closed-form method for $\Bnn$ given in 
Sec.~\ref{sec:ICASSP}, the parameter is $\lambda_1\s$. For the other three 
methods, the parameter is $\gamma_1\s.$ From the descriptions of the methods 
given in Tabs.~\ref{tab:Rnn_num}--\ref{tab:Rdd_num} it is clear that all four 
methods have a computational complexity of $\cO(M)$ flops. The brute-force 
method for $\Bnn$ is based on a search over a two-dimensional parameter space 
\cite{Jorswieck-2008-TSP}. We get $M^2$ rate points when we sample each 
parameter in $M$ grid points. The algorithm\footnote{See the source code 
available at \url{http://urn.kb.se/resolve?urn=urn:nbn:se:liu:diva-93845}.} we use to find the boundary 
out of $L$ rate pairs is similar to the mergesort algorithm, see e.g., 
\cite{Knuth-TAOCP}, which has a worst-case complexity of $\cO(L \log L)$ 
flops. Hence, the total complexity of the brute-force comparison is $\cO(M^2 
\log M)$ flops. For $\Bdn$ and $\Bnd$, we need three parameters to describe 
all pairs of potentially PO points \cite{Ho-2011-TIT}. Hence, the search is 
over $M^3$ rate points and the total complexity is $\cO(M^3 \log M)$ flops. 
For $\Bdd$, we need four parameters according to \cite{Ho-2011-TIT}, which 
would imply a total complexity of $\cO(M^4 \log M)$ flops. On the other hand, 
we noticed in Sec.~\ref{sec:Rdd} that it is straightforward to reduce the 
number of parameters to two and the total complexity to $\cO(M^2 \log M)$ 
flops.

In Tab.~\ref{tab:Complexity}, we give the complexities for the proposed 
methods and the brute-force methods. In the numerical computations, we use 
$M=500$ grid points and the tolerance $\epsilon = 5.0 \!\cdot\! 10^{-5}$.
The complexity for computing the channel constants $g_{ij},$ $\kappa_i,$ 
$\sigma_i^2,$ $i,j=1,2,$ and the final beamforming vectors is not included in 
this analysis. When we count the flops, we use the numbers given by 
\cite{Minka-Lightspeed}. The complexities of the proposed methods are at least 
one order-of-magnitude less than the corresponding brute-force methods. For 
$\Bdn$ and $\Bnd$, the complexity reduces more than four orders-of-magnitude. 
The complexity of our closed-form method for $\Bnn$ is about $20\%$ less than 
the complexity of the method presented in \cite{Mochaourab-2012-JSTSP}. This 
gain arises from the fact that our choice of parameterization of the 
beamforming vectors is more computationally efficient, which shows that the 
choice of parameterization is not unimportant. Also, the numerical method for 
$\Bnn$ has 2--5 times higher complexity than the closed-form method. On the 
other hand, the numerical method is more efficient for solving 
\eqref{eq:RnnObj}--\eqref{eq:RnnCon} for a specific $\gamma_1\s.$ For $\Bdd,$ 
the gain is one order-of-magnitude. Compared to the state-of-the-art 
brute-force method with four parameters \cite{Ho-2011-TIT}, which has 
complexity $\cO(M^4 \log M)$ the gain is even larger. On a desktop computer 
running Matlab, it takes about $50 \text{ ms}$ to find the boundaries of the 
four regions using the proposed methods. Using the brute-force methods, it 
takes a few hours to find the boundaries. One observation is that the 
numerical method for $\Bdd$ is less complex than the numerical method for 
$\Bnn$. This might seem counterintuitive since 
\eqref{eq:RddEl_Prob2_Obj}--\eqref{eq:RddEl_Prob2_Con2} is more involved than 
\eqref{eq:Rnn_scalar_opt}, but the reason is that we partly solve the former 
in closed form.

\begin{table}[H]
  \begin{center}
    \begin{tabular}{|l|c|c|c|c|}
      \hline 
      & Order & Fig. 2 & Fig. 3 & Fig. 4\\ \hline 
      \!\!\!\!$\Bnn$, numerical, Tab. \ref{tab:Rnn_num} \!\! & $\cO(M)$ 
      & $\!\!2.3 \!\cdot\! 10^5\!\!$ & $\!\!2.9 \!\cdot\! 10^5\!\!$ & 
      $\!\!5.1 \!\cdot\! 10^5\!\!$ \\
      \!\!\!\!$\Bnn$, closed-form, Tab. \ref{tab:Rnn_algorithm} \!\! & 
      $\cO(M)$ 
      & $\!\!1.0 \!\cdot\! 10^5\!\!$ & $\!\!1.0 \!\cdot\! 10^5\!\!$ & 
      $\!\!1.0 \!\cdot\! 10^5 \!\!$ \\
      \!\!\!\!$\Bnn$, closed-form, \cite{Mochaourab-2012-JSTSP} \!\! & 
      $\cO(M)$ 
      & $\!\!1.3 \!\cdot\! 10^5\!\!$ & $\!\!1.3 \!\cdot\! 10^5\!\!$ & 
      $\!\!1.3 \!\cdot\! 10^5\!\!$ \\
      \!\!\!\!$\Bnn$, brute-force, \eqref{eq:wnn} \!\! & $\!\!\! \cO(M^2 
      \log 
      M)\!\!\!$ 
      & $\!\!4.2 \!\cdot\! 10^6\!\!$ & $\!\!4.3 \!\cdot\! 10^6\!\!$ & 
      $\!\!4.4 \!\cdot\! 10^6\!\!$ \\ \hline
      \!\!\!\!$\Bdn$, closed-form, Tab. \ref{tab:Rdn_algorithm} 
      \!\!\!\!\! 
      & 
      $\cO(M)$ 
      & $\!\!7.0 \!\cdot\! 10^4\!\!$ & $\!\!6.7 \!\cdot\! 10^4\!\!$ & 
      $\!\!7.0 \!\cdot\! 10^4\!\!$ \\
      \!\!\!\!$\Bdn$, brute-force, \eqref{eq:w1dn}--\eqref{eq:w2dn} 
      \!\!\!\!\! 
      & 
      $\!\!\!\cO(M^3 \log M)\!\!\!$ 
      & $\!\!8.8 \!\cdot\! 10^8\!\!$ & $\!\!8.8 \!\cdot\! 10^8\!\!$ & 
      $\!\!8.8 \!\cdot\! 10^8\!\!$ \\ \hline
      \!\!\!\!$\Bnd$, closed-form, Tab. \ref{tab:Rdn_algorithm} 
      \!\!\!\!\! 
      & 
      $\cO(M)$ 
      & $\!\!7.0 \!\cdot\! 10^4\!\!$ & $\!\!6.7 \!\cdot\! 10^4\!\!$ & 
      $\!\!6.8 \!\cdot\! 10^4\!\!$ \\
      \!\!\!\!$\Bnd$, brute-force, \eqref{eq:w1dn}--\eqref{eq:w2dn} 
      \!\!\!\!\! 
      & 
      $\!\!\!\cO(M^3 \log M)\!\!\!$ 
      & $\!\!8.8 \!\cdot\! 10^8\!\!$ & $\!\!8.8 \!\cdot\! 10^8\!\!$ & 
      $\!\!8.8 \!\cdot\! 10^8\!\!$ \\ \hline
      \!\!\!\!$\Bdd$, numerical, Tab. \ref{tab:Rdd_num} \!\! & $\cO(M)$ 
      & $\!\!8.8 \!\cdot\! 10^4\!\!$ & $\!\!2.2 \!\cdot\! 10^5\!\!$ & 
      $\!\!1.2 \!\cdot\! 10^5\!\!$ \\
      \!\!\!\!$\Bdd$, brute-force, \eqref{eq:widd} \!\! & $\!\!\!\cO(M^2 
      \log M)\!\!\!$ 
      & $\!\!2.2 \!\cdot\! 10^6\!\!$ & $\!\!2.2 \!\cdot\! 10^6\!\!$ &  
      $\!\!2.0 \!\cdot\! 10^6\!\!$ \\ \hline	
    \end{tabular}
  \end{center}
  \caption{Computational complexity in flops for the examples in 
    Figs.~\ref{fig:sym_LowCorr}--\ref{fig:sym_DiffCorr} for the proposed 
    and brute-force methods with $M=500.$ }
  \label{tab:Complexity}
\end{table}

\section{Conclusion}
\label{sec:conclusion}
We proposed an efficient method to compute the Pareto boundary of the rate 
region for the two-user MISO IC with SIC-capable RXs. The merit of the 
proposed method, compared to the state-of-the-art, is that it avoids the 
brute-force search over all potentially PO beamforming vector pairs. The 
complexity of the proposed method is constant with respect to the number of 
transmit antennas. More importantly, we observed that the complexity gain of 
the proposed methods is a few orders-of-magnitude compared to the 
state-of-the-art brute-force methods. We achieved this by solving the 
quasi-concave optimization either by solving a cubic equation or performing a 
scalar line search. Finally, the numerical results illustrate that SIC should 
be performed when the cost of boosting the interference is small, i.e., when 
the crosstalk channel is strong or the spatial correlation of the forward and 
crosstalk channels is large.

It appears unlikely that there is any structure left in the problem that we 
can exploit in order to further improve the efficiency. Unfortunately, it 
seems that the proposed methods are not directly applicable for the general 
$K$-user MISO IC, where the number of parameters grows as $K(K-1)$ for $\Rnn$ 
\cite{Mochaourab-2011-TSP} and probably even faster for the other regions. The 
number of regions, corresponding to all possible decoding orders grows at 
least as $\cO(((K-1)!)^K)$. This number follows from the case where each 
receiver decodes all $(K-1)$ interfering signals, which can be done in 
$(K-1)!$ different orders.

The practical usefulness of our methods has been demonstrated by studies by 
others. For example, the closed-form method for computing $\Bnn$ was used in 
\cite{Litjens-2013-VTC} for a system-level assessment of inter-operator 
spectrum sharing. Also, we can use the method for the MISO broadcast channel. 
However, for this task we have to perform an extra line search to find the 
optimal power allocation. Perhaps, the methodology we brought forward here can 
be applied to other problems too.

\section*{Acknowledgment}
\addcontentsline{toc}{section}{Acknowledgment}
We would like to thank the associate editor and the three anonymous 
reviewers for their valuable comments that helped us to improve this paper.

\ifCLASSOPTIONcaptionsoff
  \newpage
\fi

\bibliographystyle{IEEEtran}




\end{document}